\documentclass[a4paper,11pt]{article} 
\usepackage{arxivpub}
\usepackage{graphicx}

\usepackage{epsfig}
\usepackage{color} % for comments
\usepackage{amssymb}
\usepackage{amsmath}
\usepackage{doi}
\usepackage{dsfont}
\usepackage{setspace}

\usepackage{verbatim}

\usepackage{slashed}% for missing E_T or missing p_T
\newcommand{\mr}{\mathrm}

\newcommand{\hjj}{$H+2~\text{jets}$}

\newcommand{\muf}{\mu_\mr{F}}
\newcommand{\mur}{\mu_\mr{R}}

\newcommand{\HAWK}{{\tt{HAWK}}}
\newcommand{\MCFM}{{\tt{MCFM}}}

\newcommand{\VBFNLO}{{\tt{VBFNLO}}}

\newcommand{\MGAMCNLO}{{\tt{MadGraph5\_aMC@NLO}}} 
\newcommand{\MADGRAPH}{{\tt{MadGraph}}} 
\newcommand{\MCATNLO}{{\tt{MC@NLO}}}

\newcommand{\POWHEG}{{\tt{POWHEG}}}
\newcommand{\POWHEGBOX}{{\tt{POWHEG~BOX}}}
\newcommand{\POWHEGBOXVV}{{\tt{POWHEG~BOX~V2}}}
\newcommand{\POWHEGBOXR}{{\tt{POWHEG~BOX~RES}}}
\newcommand{\PYTHIA}{{\tt{PYTHIA}}}

\newcommand{\PYTHIAE}{{\tt{PYTHIA8}}}
\newcommand{\HERWIG}{{\tt{HERWIG}}}
\newcommand{\HERWIGS}{{\tt{HERWIG7}}}

\newcommand{\LHE}{{\tt{LHE}}}

\newcommand{\RECOLA}{{\tt{RECOLA}}}
\newcommand{\RECOLAVV}{{\tt{RECOLA2}}}
\newcommand{\COLLIER}{{\tt{COLLIER}}}

\newcommand{\beq}{\begin{equation}}
\newcommand{\eeq}{\end{equation}}

\newcommand{\bea}{\begin{eqnarray}}
\newcommand{\eea}{\end{eqnarray}}

\newcommand{\gev}{\mr{GeV}}

\newcommand{\ptj}{p_{T,\mr{jet}}}
\newcommand{\yj}{y_{\mr{jet}}}
\newcommand{\yjstar}{y_\mr{jet_3}^{\star}}

\title{Electroweak corrections and shower effects  
to Higgs production in association with two jets at the LHC}

\author[]{Barbara J\"ager,}
\author[]{Johannes Scheller}

\affiliation[]{Institute for Theoretical Physics, University of T\"ubingen,\\ 
Auf der Morgenstelle 14, 72076 T\"ubingen, Germany}

\emailAdd{barbara.jaeger@itp.uni-tuebingen.de}
\emailAdd{johannes.scheller@uni-tuebingen.de}

\abstract{
We present an implementation of the full electroweak \hjj{} production process at hadron colliders in the framework of the \POWHEGBOX, a public tool for the matching of fixed-order perturbative calculations with parton shower generators. Our implementation allows for the simultaneous description of vector-boson fusion and Higgsstrahlung contributions. 
NLO-QCD and electroweak corrections are taken into account and matched to QCD and QED showers, respectively.
The size of the fixed-order QCD and electroweak corrections is found to be moderate, but dependent on the  considered selection cuts.  QCD shower effects slightly modify the NLO-QCD predictions and are most pronounced for distributions of non-tagging jets. The impact of QED shower effects is small.
}

\arxivnumber{JHEP09(2022)191}
\begin{document}
\maketitle
\flushbottom
\section{Introduction}
After the discovery of the Higgs boson~\cite{Chatrchyan:2012xdj,Aad:2012tfa} exploring its properties has become one of the cornerstones of the physics program of the CERN Large Hadron Collider (LHC). A particularly clean environment for probing the Higgs boson is constituted by the vector boson fusion (VBF) process, where two quarks or anti-quarks scatter via the $t$-channel exchange of an electroweak (EW) $W^\pm$ or $Z$ boson that in turn emits a Higgs boson. The quarks give rise to two jets that typically end up in the forward and backward regions of the detector with significant separation in invariant mass and rapidity, so-called ``tagging jets''. These features can be exploited for the design of selection cuts to distinguish the VBF process from QCD-induced \hjj{} production~\cite{Rainwater:1999sd}, but also from the Higgs-strahlungs process, $pp\to HV, V\to 2~\mr{jets}$ ($V=W^\pm, Z$) which is of the same order in the EW coupling as the VBF-induced \hjj{} process. 
Dedicated VBF cuts serve to efficiently suppress $HV$ contributions to the full \hjj{} final state~\cite{Ciccolini:2007ec}. However, in a more inclusive experimental setup they can become numerically relevant. Moreover, when EW corrections are considered, a separation of the EW \hjj{} production process into VBF and $HV$ contributions is no longer straightforward. 
Additionally, a detailed investigation of the $HV$ final state can be of interest itself. %
A comprehensive simulation of electroweak \hjj{} production including EW corrections therefore requires taking both topologies into account at the same time.

Precision calculations for VBF induced \hjj{} production at the next-to-leading order (NLO) in QCD have first been presented in \cite{Han:1992hr} in a so-called structure-function approach where VBF is considered as a combination of two deep-inelastic scattering processes. Subsequently, more exclusive calculations became available \cite{Figy:2003nv,Berger:2004pca} providing full information on the kinematics of the Higgs boson and the tagging jets. These calculations have been embedded in the public Monte-Carlo generators \VBFNLO{}~\cite{Arnold:2008rz} and \MCFM~\cite{Campbell:2002tg}. 
First steps towards next-to-next-to-leading order (NNLO) QCD predictions included the calculation of NLO-QCD corrections to VBF-induced $H+3~\mr{jets}$ production~\cite{Figy:2007kv} which contains the double-real and one-loop single real corrections for the $H+2~\mr{jets}$ final state, and the fully inclusive calculation of VBF-induced Higgs production in the structure-function approach \cite{Bolzoni:2010xr,Bolzoni:2011cu}. Using the novel ``projection-to-Born'' technique, in \cite{Cacciari:2015jma} fully differential NNLO-QCD results for VBF-induced Higgs production were obtained. This calculation was later on extended to three-loop accuracy in \cite{Dreyer:2016oyx}. An alternative calculation of the NNLO-QCD corrections was presented in \cite{Cruz-Martinez:2018rod}. 
NNLO-QCD corrections were found to be rather mild for inclusive quantities, but can be more pronounced when dedicated selection cuts are applied (see, for instance, the discussion in \cite{Buckley:2021gfw}).
A multijet merging for electroweak Higgs production in association with up to four jets was presented in~\cite{Chen:2021phj}. 
The impact of the ``VBF approximation'' was quantified by explicit calculations of the non-factorizable QCD corrections~\cite{Liu:2019tuy,Dreyer:2020urf} that have been neglected in previous calculations. Soft QCD effects were explored in~\cite{Bittrich:2021ztq}.

For on-shell $HV$ production NNLO-QCD corrections to inclusive cross sections are known for some time. NNLO-QCD corrections to $HW$ and $HZ$ production have been considered in~\cite{Brein:2003wg,Brein:2011vx} and implemented together with NLO-EW corrections in the numerical program {\tt vh@nnlo}~\cite{Brein:2012ne}. Further contributions to $HV$ production emerging at the second order in the strong coupling, $\alpha_s^2$, have been computed in~\cite{PhysRevD.42.2253}. Approximate results exist for contributions of gluon-initiated channels up to order $\alpha_s^3$~\cite{Altenkamp:2012sx}. 
Differential results for $HV$ production including the decay of an off-shell $V$~boson became available at NNLO-QCD accuracy in~\cite{Ferrera:2011bk,Ferrera:2014lca,Campbell:2016jau,Ferrera:2017zex}. 

%%%%%
%
Electroweak corrections to $HV$ production have been computed in~\cite{Ciccolini:2003jy,Denner:2011id}, and to VBF in~\cite{Ciccolini:2007jr,Ciccolini:2007ec,Figy:2010ct}. The public parton-level Monte-Carlo generator \HAWK~\cite{Denner:2014cla} contains the NLO-QCD and EW corrections to the full EW \hjj{} production process. In contrast to QCD mediated production processes, which typically feature EW corrections much smaller than the dominant strong corrections, 
the EW \hjj{} production process exhibits sizable EW corrections already at the level of inclusive cross sections at LHC energies $\sqrt{S}$ of up to 14~TeV. Due to Sudakov enhancements~\cite{Denner:2000jv} EW corrections are becoming even more pronounced at higher energies (relevant, for instance, at future collider facilities operating at energies far beyond the current LHC reach) and, already at moderate values of $\sqrt{S}$, in the tails of some kinematic distributions. Precision analyses thus clearly call for taking EW corrections into account at the same footing as NLO-QCD corrections. 

 Complementary to fixed-order perturbative corrections the impact of parton shower (PS) and non-perturbative effects, such as underlying event, multi-parton interactions, hadronization, should be considered. Multi-purpose Monte-Carlo generators like \PYTHIA~\cite{Sjostrand:2006za,Sjostrand:2014zea}, \HERWIG~\cite{Corcella:2000bw,Bellm:2015jjp}, or {\tt SHERPA}~\cite{Gleisberg:2008ta} provide various options to that account. However, combining fixed-order calculations with parton showers in a meaningful way requires the application of a matching formalism such as \MCATNLO~\cite{Frixione:2002ik} or \POWHEG~\cite{Nason:2004rx,Frixione:2007vw}. These were developed originally to allow for a combination of NLO-QCD calculations with QCD parton showers, but later on were applied to the matching of NLO-EW calculations with QED showers as well. 
In the context of the \POWHEGBOX~\cite{Alioli:2010xd}, a framework for the matching of fixed-order calculations with parton shower programs, VBF was one of the first processes to be implemented at NLO-QCD+PS accuracy~\cite{Nason:2009ai}. Subsequently, NLO-QCD+PS implementations of VBF~\cite{Frixione:2013mta,Campanario:2013fsa} in the \MGAMCNLO~\cite{Alwall:2014hca} and \HERWIGS~\cite{Bellm:2015jjp} generators were published, and their predictions and intrinsic uncertainties systematically compared~\cite{Jager:2020hkz}. In ref.~\cite{Luisoni:2013cuh} a \POWHEGBOX{} generator at NLO-QCD+PS accuracy for $HV$ production and $HV+\mr{jet}$ production was presented. QCD and EW corrections to the $HV$ production process including leptonic decays of the gauge boson and shower effects were provided in~\cite{Granata:2017iod}.

In this work, we wish to follow up on that count. We are presenting the implementation of an NLO+PS calculation for EW \hjj{} production at hadron colliders in the context of the \POWHEGBOX. Our calculation accounts for NLO-QCD and EW corrections to the \hjj{} final state, including VBF and $HV$ topologies at the same time. The fixed-order calculation can be combined with either QCD or QED shower effects following the \POWHEG{} prescription. 
This implementation constitutes the first dedicated Monte Carlo program for computing the NLO-EW corrections to the full EW \hjj{} final state matched with QED showers. Additionally, it provides the NLO-QCD corrections matched with QCD parton showers in the same framework. In the \POWHEGBOX{} framework up to now only separate implementations for the VBF and the $HV$ production modes existed.

This article is structured as follows: In section~\ref{sec:implementation} we describe the Monte-Carlo program we developed. Using this program, we present a detailed numerical analysis of EW \hjj{} production at the LHC in section~\ref{sec:pheno}. We conclude in section~\ref{sec:conclusions}. 

%
%=================================================

%=================================================
%
\section{Details of the implementation}
\label{sec:implementation}
In order to develop a Monte-Carlo program for EW \hjj{} production at hadron colliders accounting for NLO-QCD and EW corrections and their matching to QCD and QED showers, respectively, we resort to the \POWHEGBOXR~\cite{Jezo:2015aia}. This framework is a version of the \POWHEGBOX{} that allows for a numerically stable simulation of processes with genuinely different resonance structures, as constituted by the VBF and $HV$ topologies of the \hjj{} production process. 

Similarly to the {\tt V2}~version of the \POWHEGBOX{}, the {\tt RES}~version requires the developer to provide process-specific building blocks for the list of contributing partonic sub-processes, matrix elements at Born level, virtual and real-emission corrections, and spin- and color-correlated amplitudes for the preparation of infrared subtraction terms in the context of the FKS subtraction procedure~\cite{Frixione:1995ms}. Originally designed for NLO-QCD calculations, newer versions of the \POWHEGBOX{} additionally contain all the relevant features for NLO-EW calculations including an infrared subtraction procedure for photonic corrections~\cite{Barze:2012tt,Barze:2013fru}, and can thus be used for the implementation of EW corrections to the \hjj{} production process.

In contrast to the \POWHEGBOXVV{}, the {\tt RES} version automatically provides a phase space parameterization using a multi-channel approach which allows for an efficient sampling of different topologies contributing to the considered process.
The relevant matrix elements are extracted from the automated amplitude provider \RECOLA~\cite{Actis:2016mpe,Denner:2017vms,Denner:2017wsf}.  
We note that implementations of  \RECOLA{} in  the \POWHEGBOXR{} have been presented for diboson production and  same-sign $W$~boson scattering in refs.~\cite{Chiesa:2020ttl} and \cite{Chiesa:2019ulk}, respectively. 
For our implementation of EW \hjj{} production, we are using version~2.2.2 of this program, henceforth dubbed  \RECOLAVV. This program can be used for the extraction of the Born and real-emission amplitudes squared, the interference of the virtual (one-loop) with the Born matrix elements as well as color- and spin correlated amplitudes. It generates the building blocks on the fly and does not provide source code for standalone calculations. \RECOLAVV{} makes use of an improved memory management and is able to exploit crossing symmetries among various subprocesses. This is particularly useful in the case of EW \hjj{} production, where a priori numerous different flavor structures contribute which in turn can be reduced to only a few genuine amplitudes by employing suitable crossing relations. This is automatically taken care of by \RECOLAVV{} during process generation. Within \RECOLA{}, tensor integrals are evaluated with the help of the \COLLIER{} program library~\cite{Denner:2016kdg}. For the treatment of ultraviolet and infrared singularities, \RECOLA{} uses dimensional regularization per default. 

Our \POWHEGBOX{} implementation allows to generate event files in the \LHE{} format~\cite{Alwall:2006yp} and to interface them with the \PYTHIA{} parton shower program. 
Users can select a mode to either match the NLO-EW calculations with a QED shower, or the NLO-QCD calculations with a QCD shower. 
While, naively, the \POWHEGBOX{} could also be interfaced to \PYTHIA{} by starting the shower evolution at the \POWHEG{} scale, small differences in the evolution scales used by the two programs could possibly lead to undercounting or double counting of some phase space regions and thus spoil the correctness of the calculation. We therefore provide an interface that relies on the {\tt PowhegHook} of \PYTHIAE{} to guarantee the matching of the QCD shower to the \POWHEGBOX{} calculation in a consistent way. This plug-in starts the shower evolution at the kinematic limit and subsequently vetoes any emission harder than the one generated by the \POWHEGBOX{}. If the QED shower is turned on, we perform a similar customized matching procedure. 

To validate our implementation we performed a variety of checks: 
\begin{itemize}
\item 
We prepared two versions of our \POWHEGBOX{} implementation. The default one is based on amplitudes provided by \RECOLAVV. An alternative version resorts to tree-level amplitudes prepared with a semi-automated tool based on \MADGRAPH~\cite{Alwall:2007st} within the \POWHEGBOX{} machinery. We checked that at the level of individual phase-space points all amplitudes of these two implementations agree within double-precision accuracy. 

\item At LO as well as NLO-QCD and NLO-EW accuracy we performed a comparison of the results obtained with our default implementation with results obtained with the \HAWK{} generator~\cite{Denner:2014cla,Ciccolini:2007jr,Ciccolini:2007ec,Denner:2011id,Denner:2018opp}, finding full agreement at the level of cross sections and differential distributions. For the sake of this comparison, in \HAWK{} we implemented additional differential distributions sensitive to a third jet and to real photon radiation, again finding full agreement with our calculation.

\end{itemize}

%=================================================

%=================================================
%
\section{Phenomenological results}
\label{sec:pheno}
In this section we wish to address two major issues: First, we will discuss how the PS manifests itself differently in the two kinematic regions typical for $HV$ and for VBF analyses which can both be properly simulated with our \POWHEGBOX{} implementation of EW \hjj{} production. 
Second, we will explore the impact of QED shower effects on top of EW corrections.

For our numerical studies we consider proton-proton collisions at the LHC with a center-of-mass energy of $\sqrt{S}=13$~TeV. We use the {\tt NNPDF3.1luxQED-NLO} set of parton distributions functions (PDFs), corresponding to the identifier $324900$ in the {\tt LHAPDF6} library~\cite{Buckley:2014ana}, and the associated strong coupling with $\alpha_s(M_Z)=0.118$. Jets are reconstructed according to the anti-$k_T$ algorithm~\cite{Cacciari:2008gp} with an $R$-parameter of 0.4 with the help of the {\tt FASTJET} package~\cite{Cacciari:2011ma}. With EW corrections turned on, photons appear in the final state. While we do not consider any explicit photon distributions and the jet algorithm is only clustering color-charged partons in our analysis, photons may still enter implicitly through the dressing of the jet: If a photon is separated from a jet by less than $\Delta{}R_{j\gamma}=0.1$ in the rapidity-azimuthal angle plane, the two objects are recombined. 

For EW parameters we use the $G_\mu$ scheme, fixing as input the values of the Fermi constant, $G_\mu=1.16637\times 10^{-5}\gev^{-2}$, and the masses of the $W$ and $Z$ bosons. Other EW parameters such as the weak mixing angle and the EW coupling $\alpha$ are computed thereof via tree-level relations. Throughout we are using the complex-mass scheme. 
For the masses and widths of the electroweak gauge bosons and the Higgs boson we use pole masses and pole widths, corresponding to the on-shell masses according to the particle data group (PDG)~\cite{Zyla:2020zbs}: 
\bea
m_W &=& 80.379~\gev\,,\quad
\Gamma_W = 2.085~\gev\,,\quad
\nonumber \\
m_Z &=& 91.1876~\gev\,,\quad
\Gamma_Z = 2.4952~\gev\,,\quad
\nonumber \\
m_H &=& 125.25~\gev\,,\quad
\Gamma_H = 3.2\times10^{-3}~\gev\,,\quad
\nonumber \\
m_t &=& 172.76~\gev\,,\quad
\Gamma_t = 1.42~\gev\,. 
\eea 
For our simulations we assume five massless external quarks. Contributions with external top quarks are disregarded throughout, whereas massive fermion loops are taken into account in the virtual EW corrections. The Cabibbo-Kobayashi-Maskawa matrix is assumed to be diagonal, i.e.\ effects of mixing between different quark generations are neglected. Initial-state photon contributions are not taken into account. In~\cite{Ciccolini:2007ec} such contributions were found to yield a correction of roughly one percent to the EW~\hjj{} production cross section, and thus to be subleading to quark-initiated channels. 
The renormalization and factorization scales, $\mur$ and $\muf$, are set dynamically by the arithmetic mean of the transverse momenta of the two outgoing partons $i_1, i_2$ in the underlying Born configuration of each event, 
\beq
\mur=\muf = \frac{p_{T,i_1}+p_{T,i_2}}{2} \,. 
\eeq

Unless stated otherwise we use the {\tt Monash~2013} tune~\cite{Skands:2014pea} of \PYTHIA{} version~8.240 for our PS simulations, dubbed \PYTHIAE{} in the following. 
For our default setup, we use the global recoil scheme of \PYTHIAE. We explicitly state if the more recent dipole recoil scheme for the space-like shower is used instead. 
Underlying event, hadronization, and multi-parton interactions are turned off. QED shower effects are switched off when we consider PS effects on the NLO-QCD results. In the context of NLO-QCD results matched to PS the acronym NLO-QCD+PS denotes the NLO-QCD calculation matched to a QCD parton shower. Contrarily, QCD shower effects are switched off when \PYTHIA{} is matched to NLO-EW results. That is, whenever we consider NLO-EW results matched to \PYTHIA's QED shower the acronym NLO-EW+PS refers to predictions at NLO-EW accuracy matched with a QED shower, but without QCD shower effects.

For our numerical studies we consider three distinct experimental scenarios: In the so-called {\em $HV$ setup} we apply cuts that favor Higgsstrahlung contributions, while in the {\em VBF setup} selection cuts typical for a VBF analysis are applied. We also consider an {\em inclusive} cut set which corresponds to only basic selection cuts.
In the inclusive setup we require the presence of at least two jets fulfilling minimal requirements on transverse momentum and rapidity,
\beq
\label{eq:inc-cuts1}
\ptj> 25~\gev\,,\quad 
|\yj|< 4.5 \,.
\eeq
The two hardest jets satisfying these requirements are called ``tagging jets''. These two jets have to satisfy an invariant mass cut of
\beq
\label{eq:inc-cuts2}
m_{jj}>60~\gev\,.
\eeq

In the VBF setup we impose the same cuts on the jets, 
\beq
\label{eq:vbf-cuts1}
\ptj>25~\gev\,,\quad
|\yj|<4.5\,.
\eeq
Again, the two hardest jets satisfying these requirements are called ``tagging jets''. To fulfill the VBF cuts, the tagging jets have to exhibit an invariant mass of
\beq
\label{eq:vbf-cuts2}
m_{jj}>600~\gev\,, 
\eeq
and be well separated in rapidity, 
\beq
\label{eq:vbf-cuts3}
\Delta y_{jj}>4.5\,.
\eeq
We also require the jets to be located in opposite hemispheres of the detector, corresponding to 
\beq
\label{eq:vbf-cuts4}
y_\mr{jet_1}\cdot y_\mr{jet_2}<0\,.
\eeq

In the $HV$ cut set, the criteria on the jets are modified to 
\beq
\label{eq:hv-cuts1}
\ptj>25~\gev\,,\quad 
|\yj|<2.5\,.
\eeq
The system of the two tagging jets has to exhibit an invariant mass in the range 
\beq
\label{eq:hv-cuts2}
60~\gev<m_{jj}<140~\gev\,.
\eeq
In each scenario, events are disregarded if they do not exhibit (at least) two jets fulfilling the criteria quoted above. The presence of additional jets has no consequence on the event selection. However, when presenting rapidity-related observables of a third jet below, we only consider subleading  jets 
fulfilling the additional requirements on transverse momentum and rapidity of 
\beq
\label{eq:jet3-cuts}
p_{T,\mr{jet_3}}>25~\gev\,,\quad 
|y_\mr{jet_3}|<4.5\,.
\eeq

We first consider NLO-QCD corrections and PS effects. Within the inclusive cuts of eqs.~(\ref{eq:inc-cuts1})--(\ref{eq:inc-cuts2}) the NLO-QCD corrections enhance the inclusive LO cross section by less than $1\%$. PS effects cause a slight reduction of the NLO-QCD cross section by about $7\%$.
%
% LO: 
%# total_INC_cuts index     45
% 0.00000000D+00 0.10000000D+01 0.32500047D+01 0.17605685D-02
%
% NLO-QCD
%# total_INC_cuts index     45
% 0.00000000D+00 0.10000000D+01 0.32715314D+01 0.23163713D-02
%
%NLO+QCD+PS: (PY8)
%# total_INC_cuts index     45
% 0.00000000E+00 0.10000000E+01 0.30456736E+01 0.16922338E-02
%
%NLO+QCD+PS: (PY8-dipole)
%# total_INC_cuts index     45
% 0.00000000E+00 0.10000000E+01 0.30001770E+01 0.16959375E-02
%
Within the VBF cuts of eqs.~(\ref{eq:vbf-cuts1})--(\ref{eq:vbf-cuts4}) the NLO-QCD corrections reduce the LO results by almost 10\%, and the PS has an additional impact of about -8\%. 
For the $HV$-specific cuts of eqs.~(\ref{eq:hv-cuts1})--(\ref{eq:hv-cuts2}) we find NLO-QCD corrections of about +29\%, and a -4\% effect of the PS. 
% LO: 
%# total_VBF_cuts index      1
% 0.00000000D+00 0.10000000D+01 0.95427524D+00 0.68209486D-03
%# total_HV__cuts index     23
% 0.00000000D+00 0.10000000D+01 0.61096487D+00 0.13845883D-02
%
% NLO QCD:
%# total_VBF_cuts index      1
% 0.00000000D+00 0.10000000D+01 0.86179896D+00 0.12562238D-02
%
%# total_HV__cuts index     23
% 0.00000000D+00 0.10000000D+01 0.79139837D+00 0.16458506D-02
%
% NLO-QCD (PY8)
%# total_VBF_cuts index      1
% 0.00000000E+00 0.10000000E+01 0.78897221E+00 0.12028787E-02
%
%# total_HV__cuts index     23
% 0.00000000E+00 0.10000000E+01 0.75805570E+00 0.11912427E-02
Let us note that these numbers depend very much on the setup and can change significantly if slightly different cuts are applied.

\begin{figure}
\includegraphics[angle=0,scale=0.5]{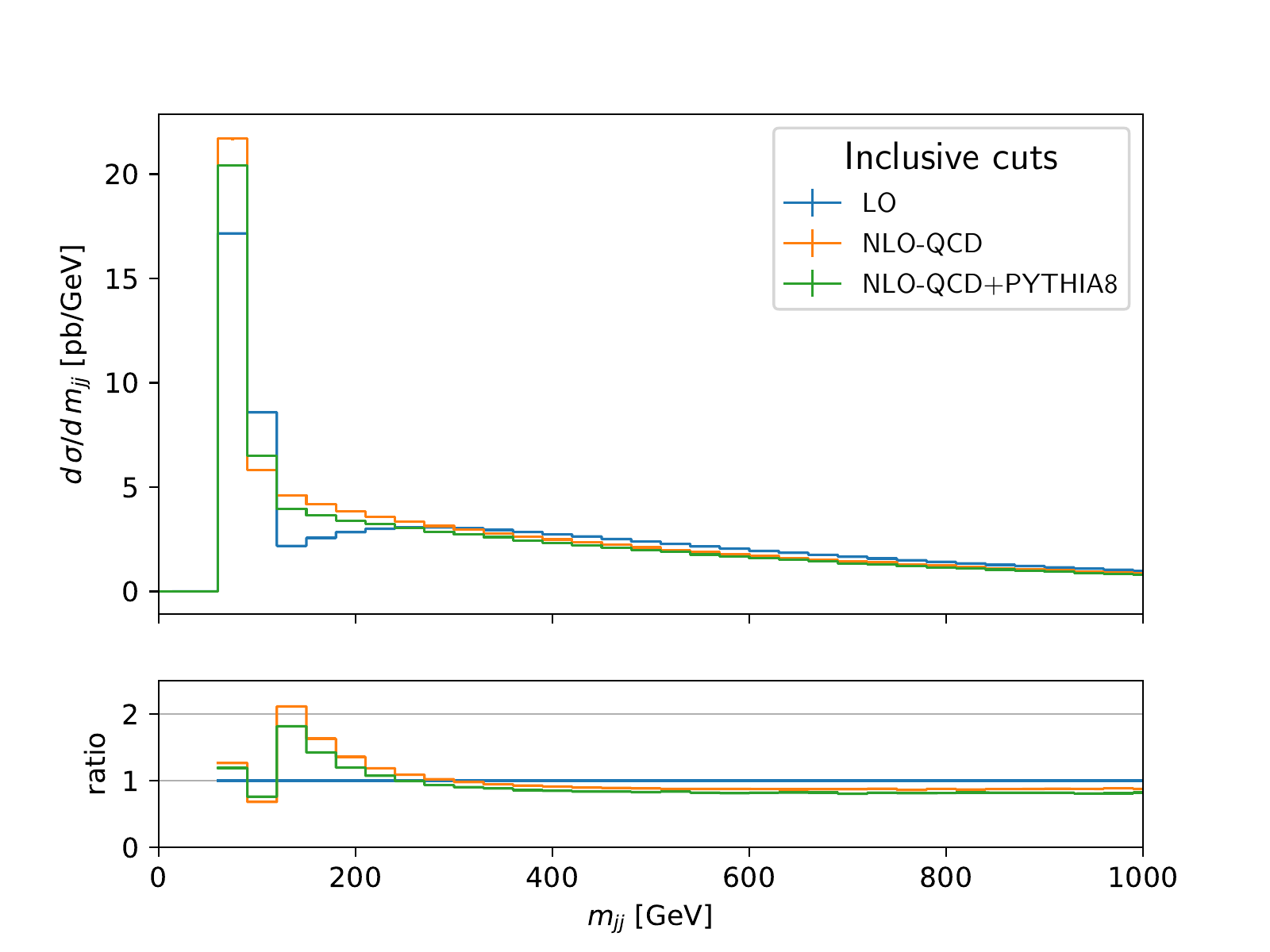}
\includegraphics[angle=0,scale=0.5]{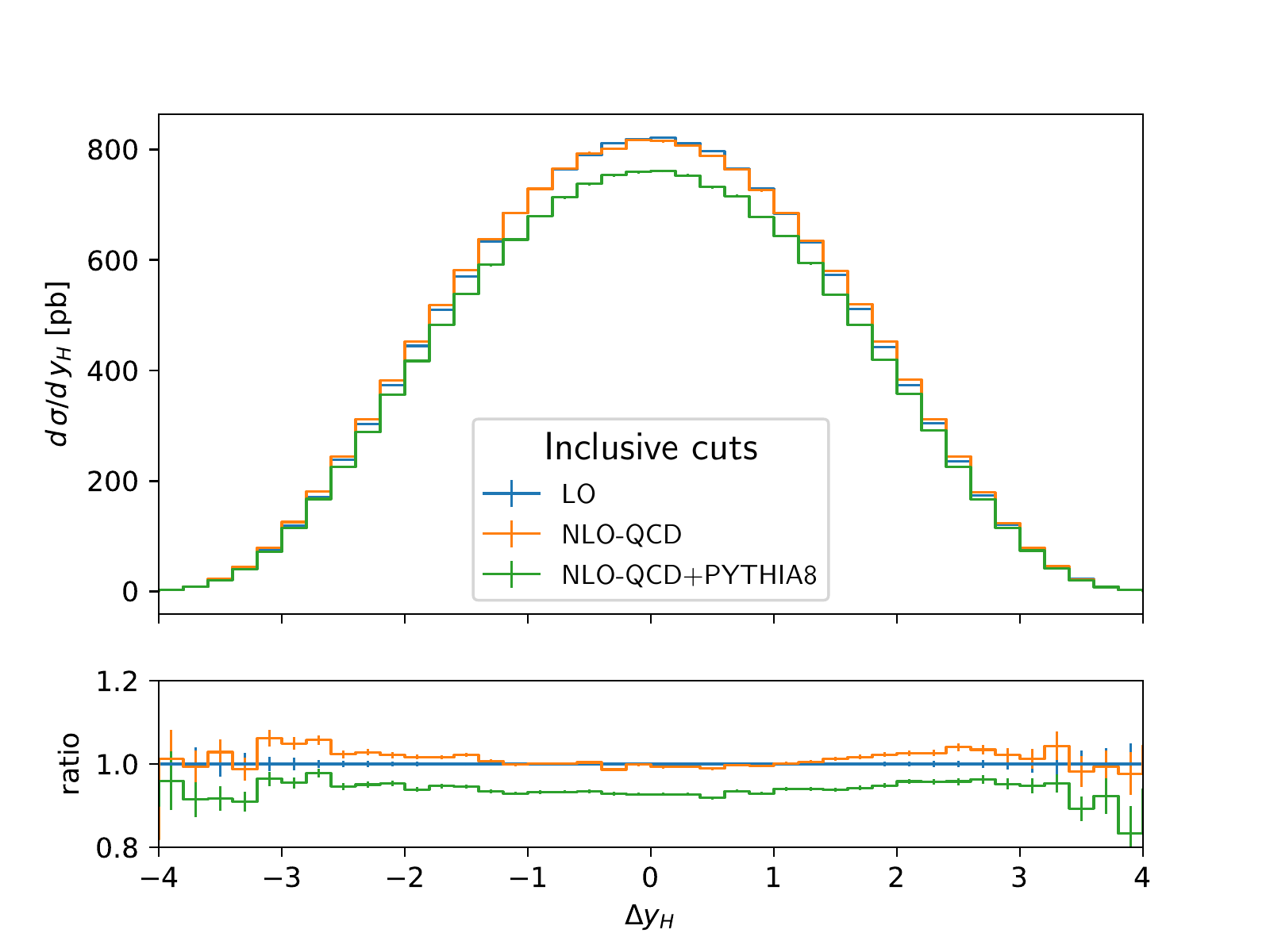}
\caption{\it 
\label{fig:qcd-inc-jets}
Invariant-mass distribution of the two tagging jets (left) and rapidity of the Higgs boson  (right) at LO (blue), NLO-QCD (red) and NLO-QCD+PS (green) accuracy within the inclusive cuts of eqs.~(\ref{eq:inc-cuts1})--(\ref{eq:inc-cuts2}). 
The ratios of the NLO-QCD to the LO (red) and of the NLO-QCD+PS to the LO results (green) are shown in the respective lower panels. 
}
\end{figure}

The impact of the QCD corrections on the shape of kinematic distributions is best illustrated by selected distributions. The r.h.s.\ of figure~\ref{fig:qcd-inc-jets} shows the invariant mass distribution of the two tagging jets.  While in the phase-space region of large $m_{jj}$ that is dominated by the VBF topology the NLO-QCD corrections cause a relatively constant increase in normalization, in the low-$m_{jj}$ region associated with $HV$ contributions the LO distribution is smeared considerably by the NLO-QCD corrections. This effect  results in a pronounced reduction of the peak associated with the dijet system stemming from the quasi on-shell decay of a massive gauge boson, and a redistribution of events towards larger values of the tagging jets' invariant mass. 
Rather uniform NLO-QCD corrections and PS effects are found instead for distributions of the individual  tagging jets and the Higgs boson, such as the rapidity of the Higgs boson shown on the l.h.s.\ of figure~\ref{fig:qcd-inc-jets}.  
%
%
%%%%
%%%%
%

\begin{figure}
\includegraphics[angle=0,scale=0.5]{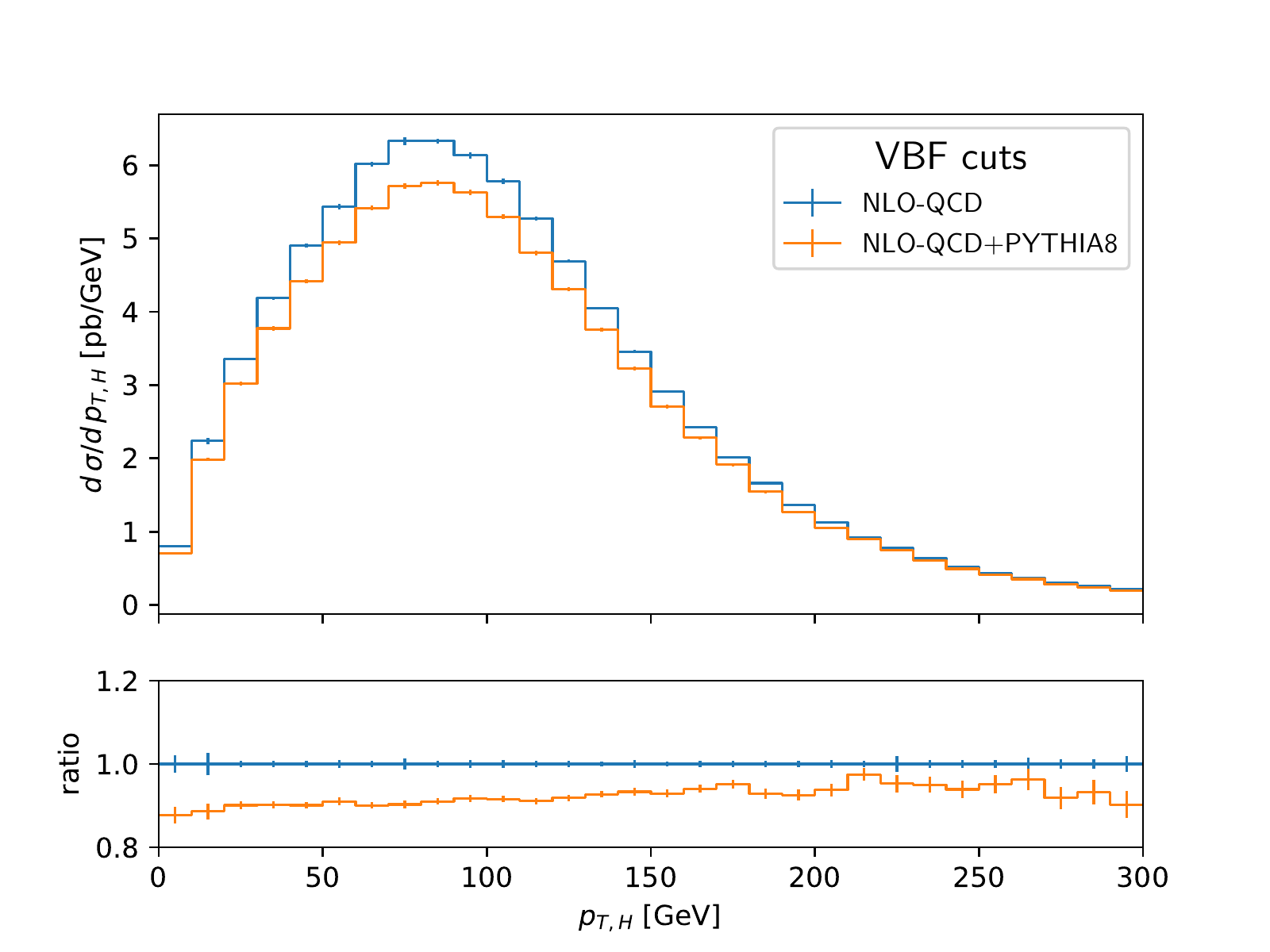}
\includegraphics[angle=0,scale=0.5]{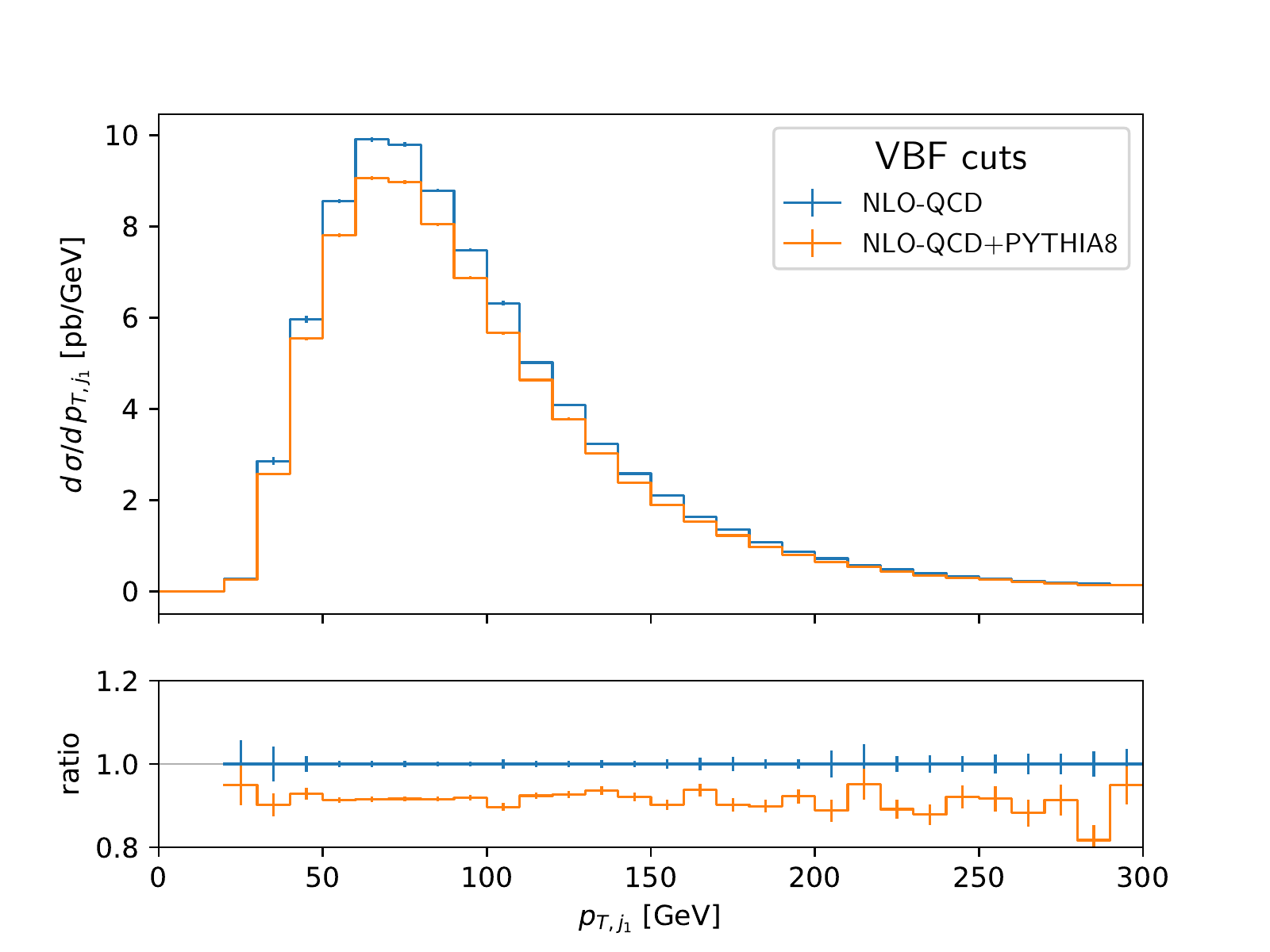}
\caption{\it 
\label{fig:qcd-vbf-pt}
Transverse-momentum distribution of the Higgs boson (left) and of the hardest tagging jet (right) at NLO-QCD (blue) and NLO-QCD+PS (red) accuracy within the VBF cuts of eqs.~(\ref{eq:vbf-cuts1})--(\ref{eq:vbf-cuts4}). 
The ratios of the NLO-QCD+PS to the NLO-QCD results are shown in the respective lower panels. 
}
\end{figure}

\begin{figure}
\includegraphics[angle=0,scale=0.5]{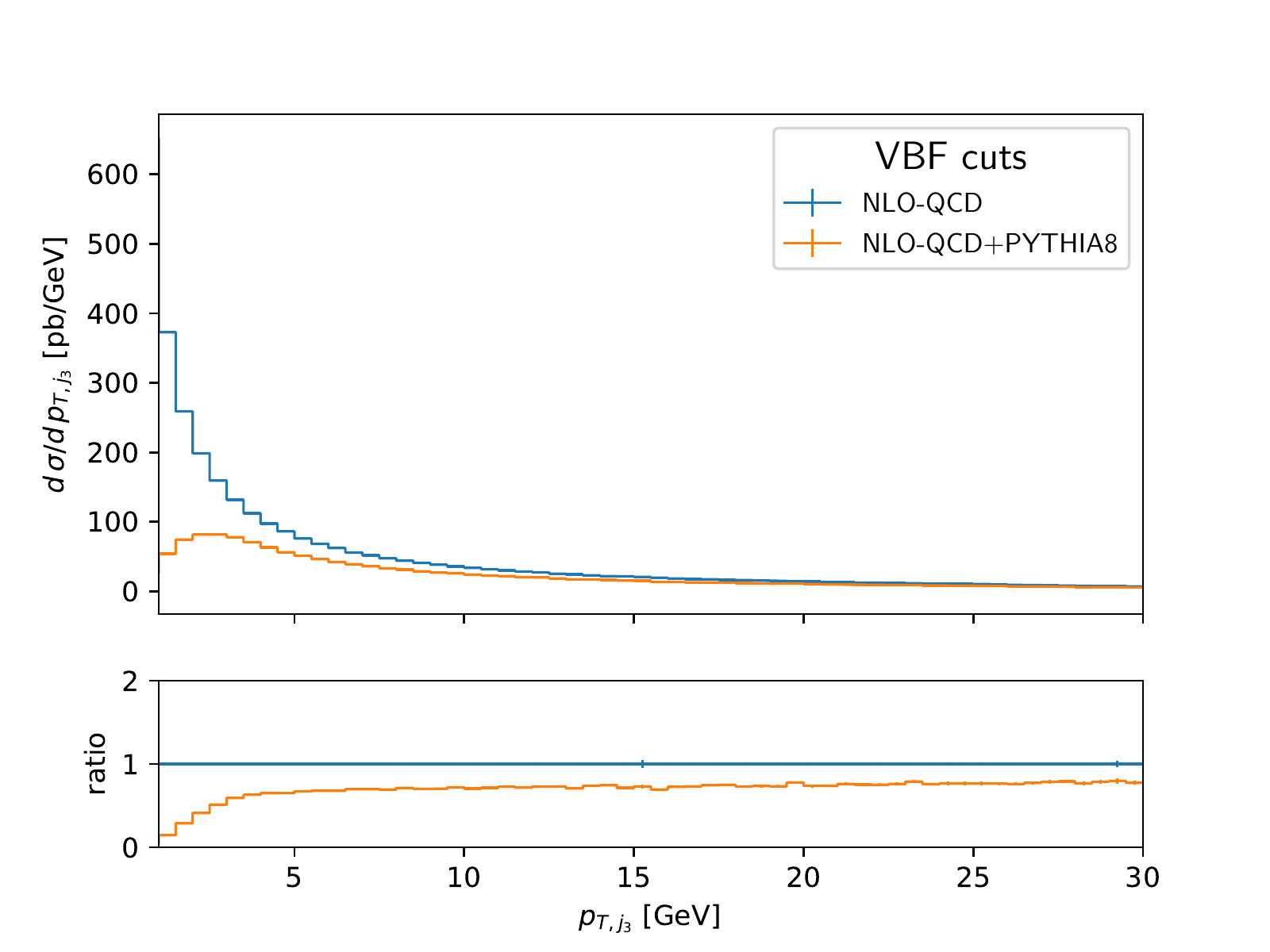}
\includegraphics[angle=0,scale=0.5]{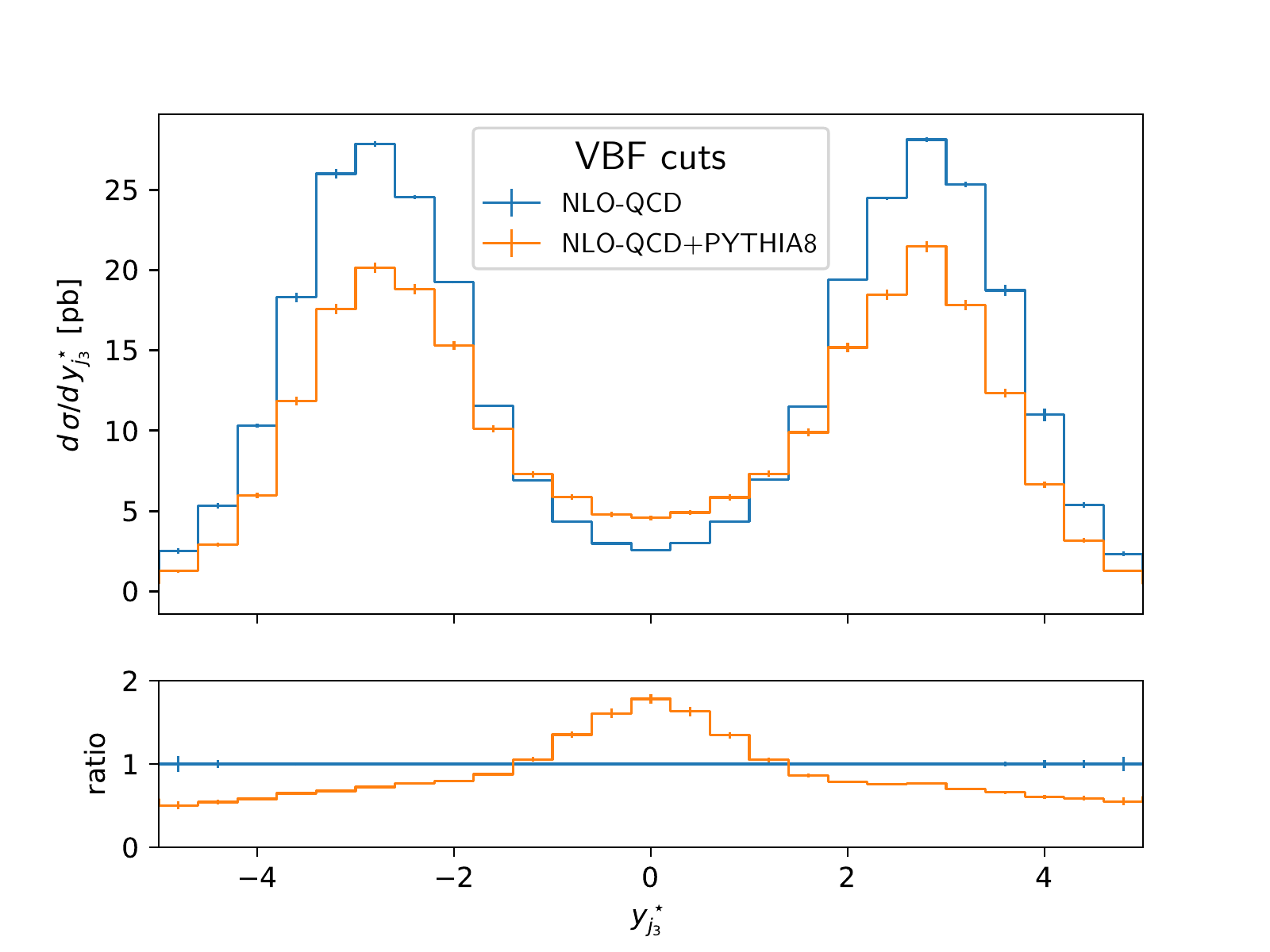}
\caption{\it 
\label{fig:qcd-vbf-jet3}
Transverse-momentum distribution (left) and $\yjstar$~variable of the third-hardest jet (right) at NLO-QCD (blue) and NLO-QCD+PS (red) accuracy within the VBF cuts of eqs.~(\ref{eq:vbf-cuts1})--(\ref{eq:vbf-cuts4}) and the extra requirements of eq.~(\ref{eq:jet3-cuts}) on the third jet. 
The ratios of the NLO-QCD+PS to the NLO-QCD results are shown in the respective lower panels.
}
\end{figure}

\begin{figure}
\includegraphics[angle=0,scale=0.5]{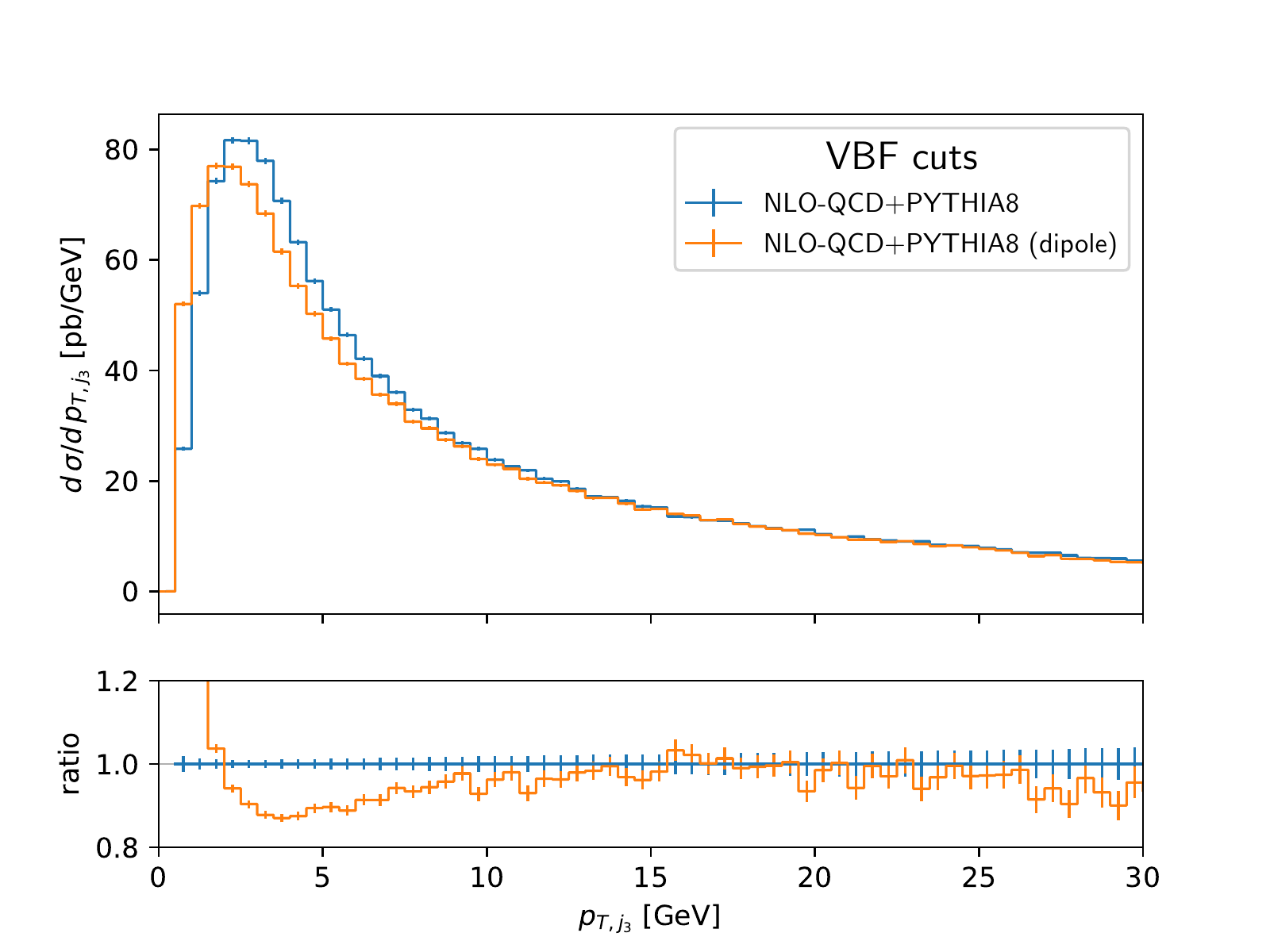}
\includegraphics[angle=0,scale=0.5]{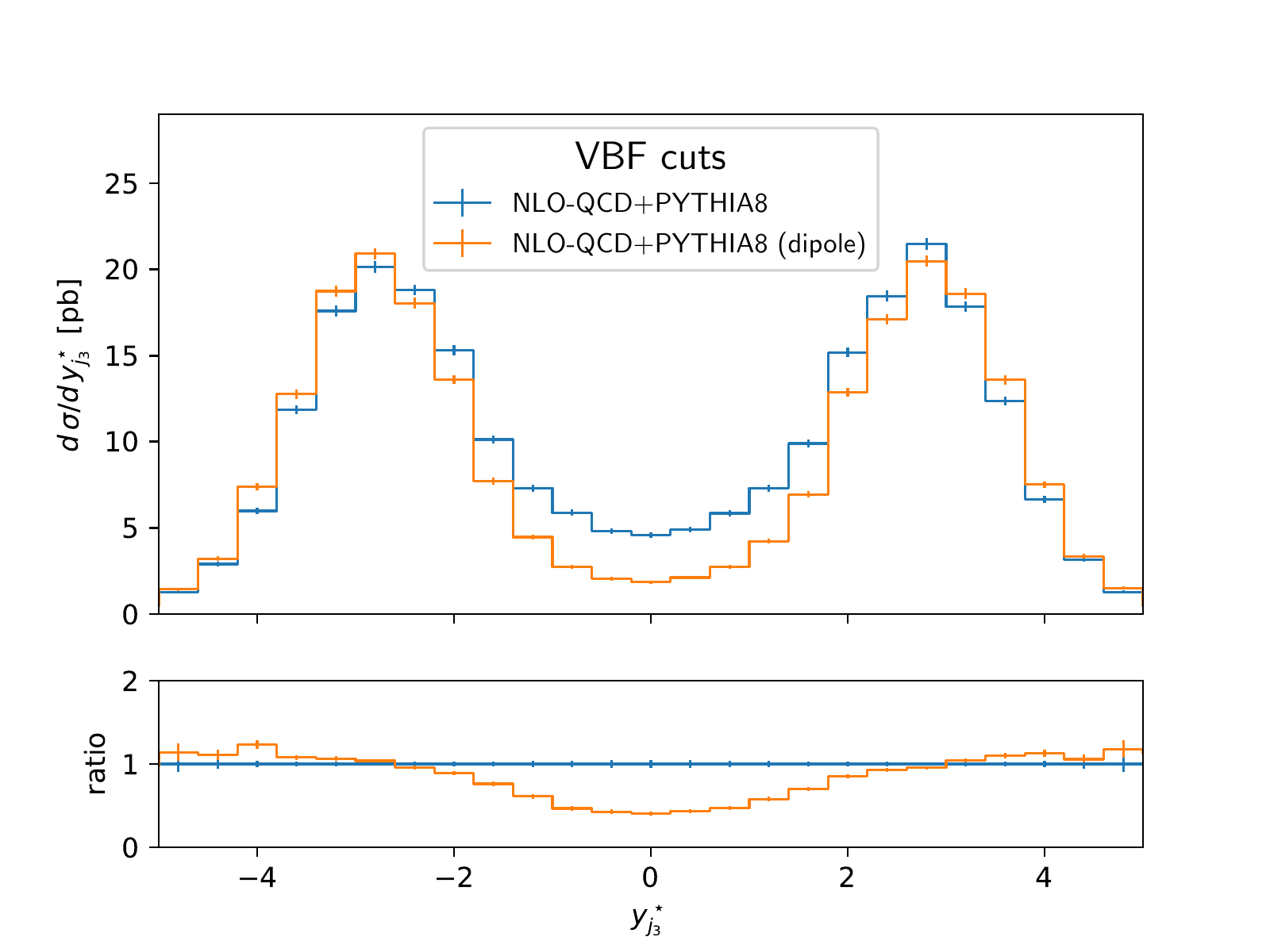}
	\caption{\it 
		\label{fig:qcd-vbf-dipole}
		Transverse-momentum distribution (left) and $\yjstar$~variable of the third-hardest jet (right) at NLO-QCD+PS accuracy with the default global recoil scheme (blue) and the dipole recoil scheme (red) of \PYTHIAE{}  within the VBF cuts of eqs.~(\ref{eq:vbf-cuts1})--(\ref{eq:vbf-cuts4}) and the extra requirements of eq.~(\ref{eq:jet3-cuts}) on the third jet.  
		The ratios of the dipole to the default shower results are shown in the respective lower panels.
		}
\end{figure}

%
In the VBF setup PS effects are typically small, as illustrated by the transverse momentum distribution of the Higgs boson and the hardest tagging jet in figure~\ref{fig:qcd-vbf-pt}. 
%
%%%%
%%%%
%
More pronounced differences between NLO-QCD and NLO-QCD+PS predictions can be observed in distributions of the third-hardest jet, depicted in figure~\ref{fig:qcd-vbf-jet3}. 
%
%%%%
%%%%
% 
In the NLO-QCD calculation, a third jet can only stem from the real-emission corrections. In the NLO-QCD+PS simulation, sub-leading jets can also be generated by the parton shower. In each of these cases, the description of observables related to non-tagging jets lacks the perturbative accuracy of observables that are already defined at Born level. Increasing the precision of predictions for the third jet would require a calculation that provides NLO-QCD corrections to $H+3~\text{jets}$, as accomplished in ref.~\cite {Figy:2007kv} for the VBF topology and in ref.~\cite{Campanario:2013fsa} for the full EW $H+3~\text{jets}$ production process.  
Indeed, we find that within our \hjj~simulation the transverse-momentum distribution of the third jet exhibits the typical increase towards low values of $\ptj$ at fixed order, which is dampened by the PS. The relative rapidity position of the third jet with respect to the two tagging jets is encoded in the $\yjstar$~distribution, defined as 
\beq
\yjstar = y_\mr{jet_3} - \frac{y_\mr{jet_1}+y_\mr{jet_2}}{2}\,.
\eeq
While at NLO-QCD the third jet barely ends up in the rapidity region between the two tagging jets, this region is filled up to some extent by the PS. It is well known (see, e.g.,  ref.~\cite{DelDuca:2006hk} for the first investigation of QCD shower effects on VBF matched with LO matrix elements, or ref.~\cite{Jager:2020hkz} for a more recent assessment of matching effects in VBF) that this feature strongly depends on the chosen PS, and can be ameliorated by the inclusion of higher-order corrections for observables of the third jet. 
According to~\cite{Cabouat:2017rzi} the unphysical enhancement of radiation by the shower in the central region is caused by the global distribution of the radiation recoil, which is not a realistic assumption for the VBF process. In the VBF approximation, color flow between the initial state quark lines is entirely disregarded.   
In~\cite{Jager:2020hkz}, it was shown that a dipole recoil shower provides a more suitable description in this case. 
Similar findings were reported in \cite{Konar:2022bgc}. 
Our results confirm that the dipole-shower option of \PYTHIAE{} suppresses unphysical radiation in the central rapidity region when VBF cuts are applied. This can be seen in figure~\ref{fig:qcd-vbf-dipole}, where we compare  the two recoil schemes for the relative rapidity position of the third jet and its transverse momentum.
%
%%%%
%%%%
%

\begin{figure}
\includegraphics[angle=0,scale=0.5]{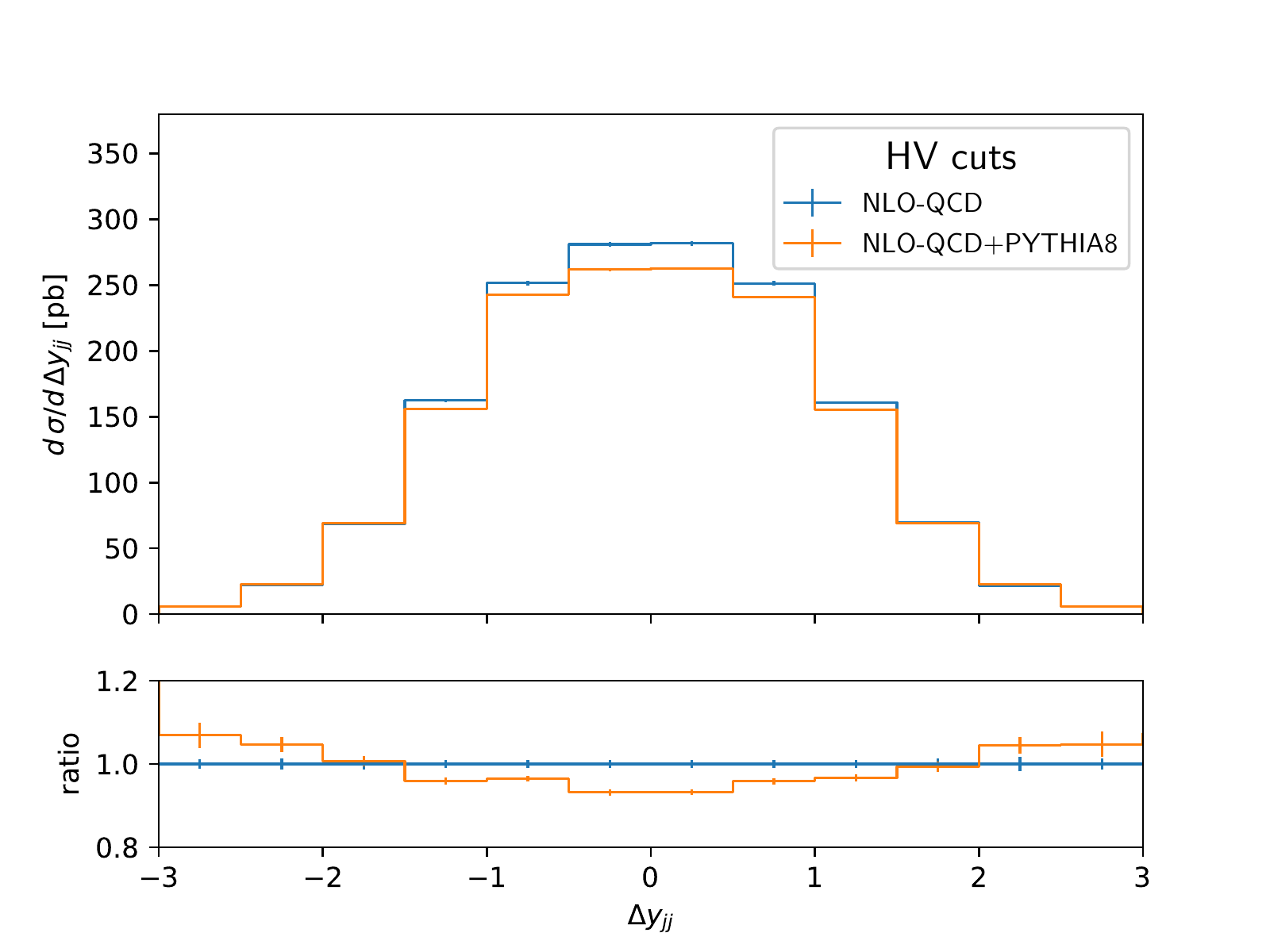}
\includegraphics[angle=0,scale=0.5]{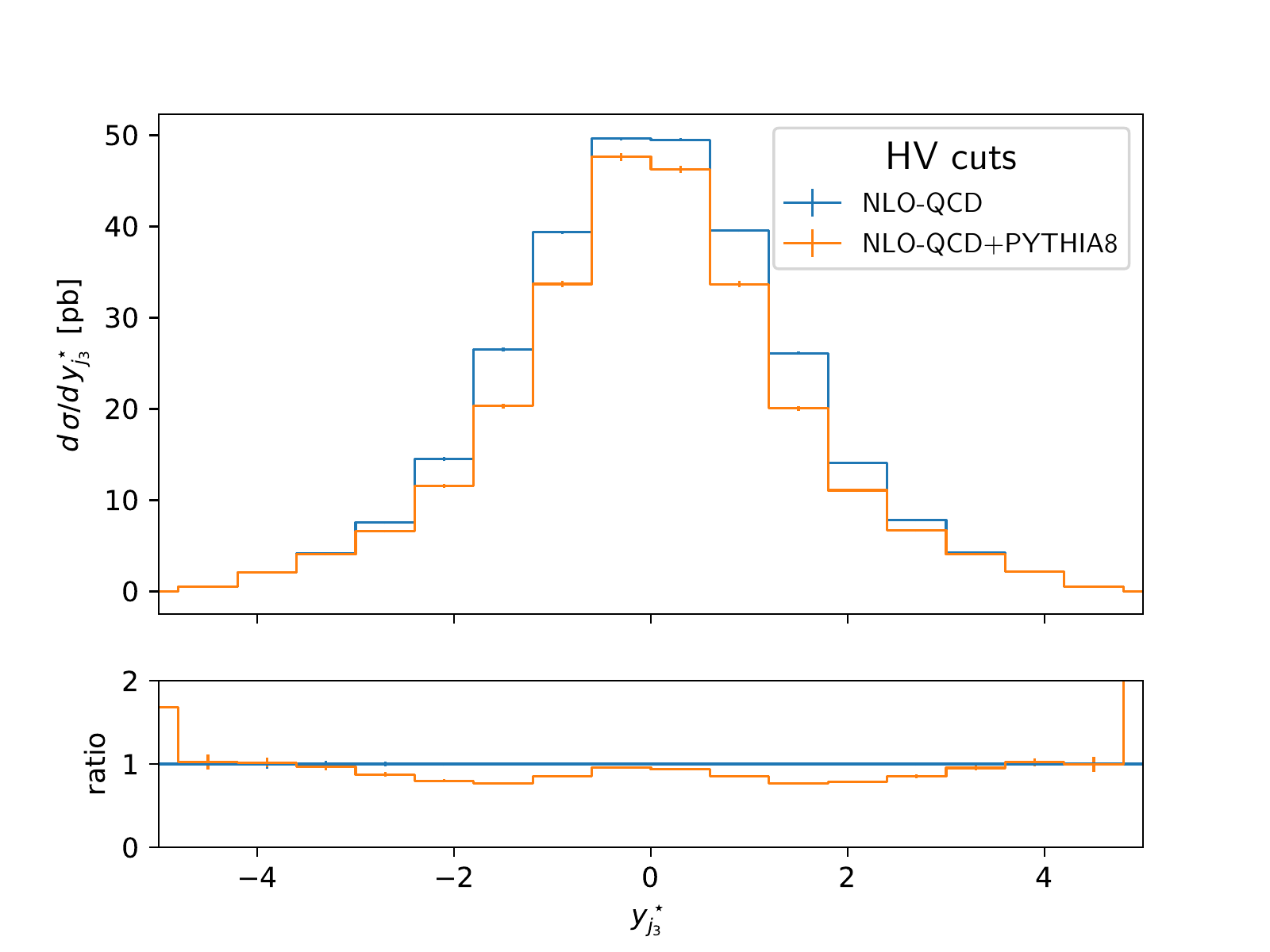}
\caption{\it 
\label{fig:qcd-hv}
Rapidity separation of the two tagging jets (left) and $\yjstar$~variable of the third-hardest jet (right) at NLO-QCD (blue) and NLO-QCD+PS (red) accuracy within the HV cuts of eqs.~(\ref{eq:hv-cuts1})--(\ref{eq:hv-cuts2}). In the latter distribution the extra requirements of eq.~(\ref{eq:jet3-cuts}) are imposed on the third jet. 
The ratios of the NLO-QCD+PS to the NLO-QCD results are shown in the respective lower panels. 
}
\end{figure}

A very different behavior of the third jet is observed in the $HV$ setup which allows for configurations where the two hardest jets are very close in invariant mass and rapidity, see figure~\ref{fig:qcd-hv}. %
%%%%
%%%%
Configurations of close tagging jets are removed by the cuts of eqs.~(\ref{eq:vbf-cuts2})--(\ref{eq:vbf-cuts3}) in the VBF setup, but constitute a large part of the $HV$ production cross section. It turns out that in the $HV$ setup the third jet does not steer clear of the rapidity region in between the two jets, but is preferentially located at central rapidities, in between the two tagging jets.

% LO: 
%# total_INC_cuts index     45
% 0.00000000D+00 0.10000000D+01 0.32500047D+01 0.17605685D-02
%
%# total_VBF_cuts index      1
% 0.00000000D+00 0.10000000D+01 0.95427524D+00 0.68209486D-03
%# total_HV__cuts index     23
% 0.00000000D+00 0.10000000D+01 0.61096487D+00 0.13845883D-02

% NLO-EW:
%# total_INC_cuts index     45
% 0.00000000D+00 0.10000000D+01 0.30519378D+01 0.87932739D-03%
%
%# total_VBF_cuts index      1
% 0.00000000D+00 0.10000000D+01 0.87076097D+00 0.37991590D-03
%
%# total_HV__cuts index     23
% 0.00000000D+00 0.10000000D+01 0.57028754D+00 0.68456665D-03

Let us now turn to a discussion of EW corrections and QED shower effects. The NLO-EW corrections modify the LO cross section within the inclusive cut set by $-6\%$, which agrees with the expectation from related work in the literature~\cite{Ciccolini:2007ec}. Similarly, NLO-EW corrections of about $-9\%$ are found for the VBF setup, and about $-7\%$ for the $HV$~setup. QED shower effects modify the NLO-EW cross sections within all considered cut scenarios only to a small extent.  
%
%
%%%%
\begin{figure}
\includegraphics[angle=0,scale=0.5]{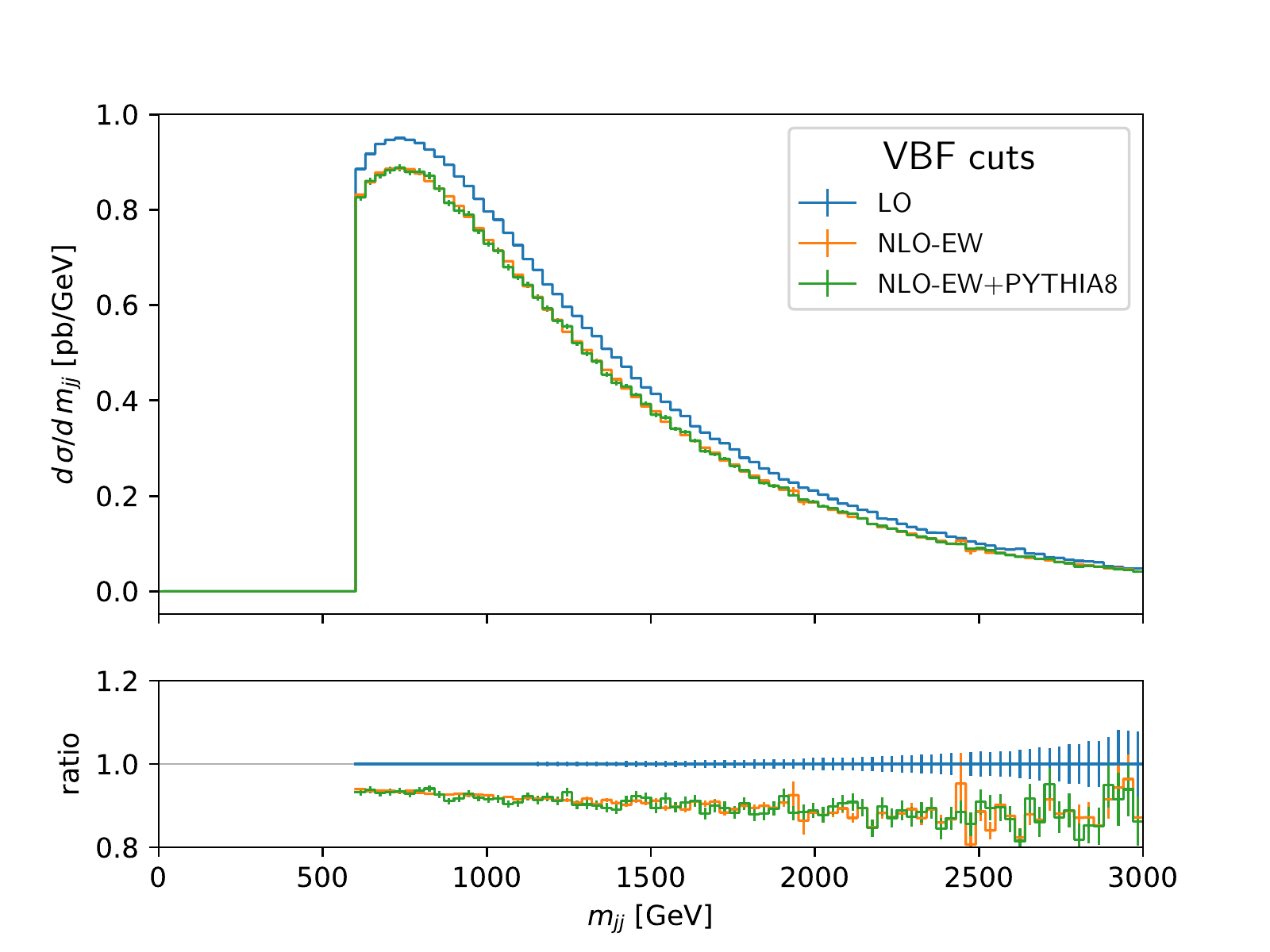}
\includegraphics[angle=0,scale=0.5]{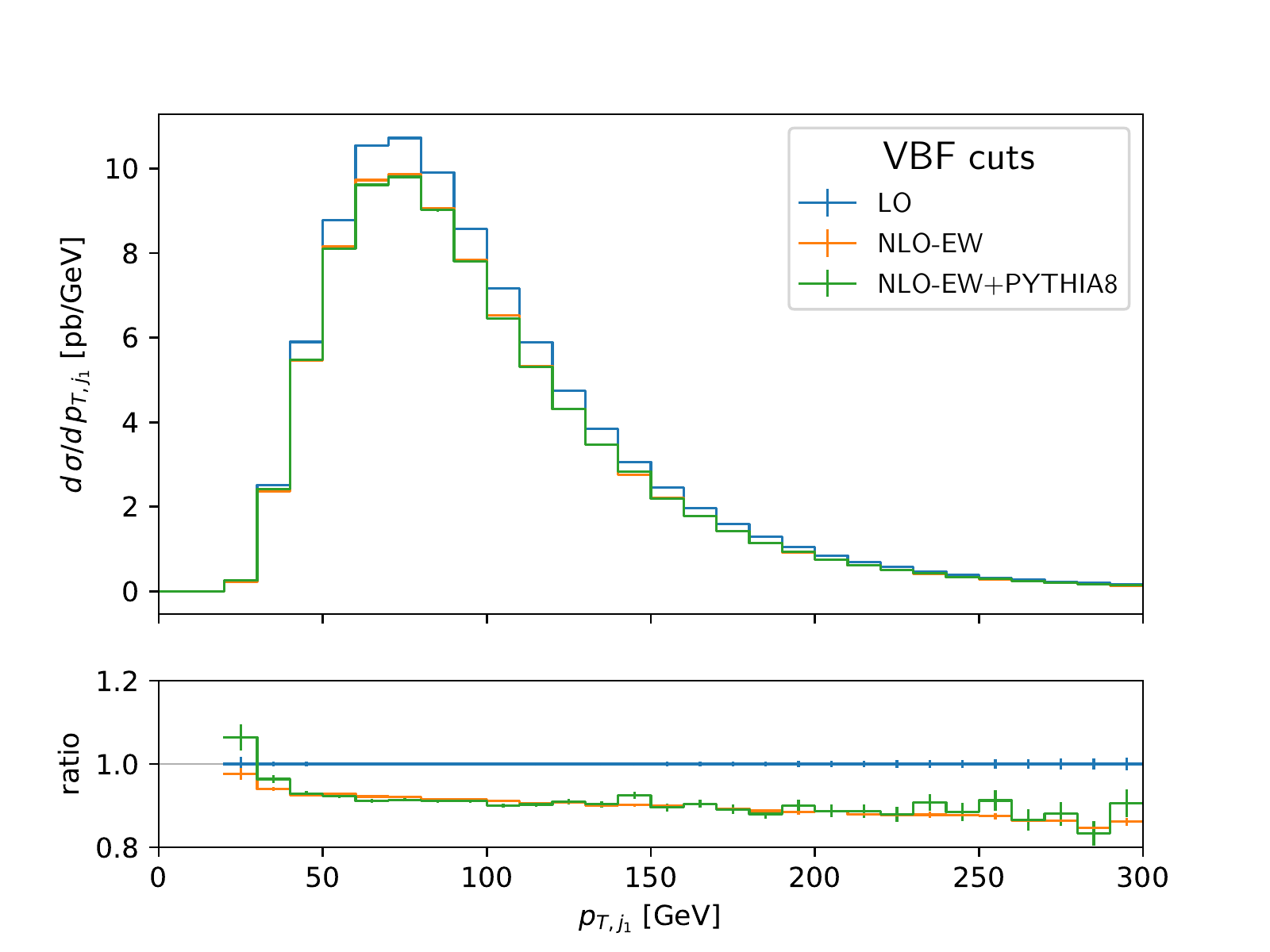}
\caption{\it 
\label{fig:ew-vbf-pt}
Invariant-mass distribution of the two tagging jets (left) and transverse-momentum distribution of the hardest tagging jet (right) at LO (blue), NLO-EW (red), and NLO-EW+PS (green) accuracy within the VBF cuts of eqs.~(\ref{eq:vbf-cuts1})--(\ref{eq:vbf-cuts4}). 
The ratios of the NLO-EW+PS and the NLO-EW to the LO results are shown in the respective lower panels. 
}
\end{figure}
%%%%
% 
Figure~\ref{fig:ew-vbf-pt} shows that for the invariant-mass distribution of the two tagging jets within the VBF cuts of eqs.~(\ref{eq:vbf-cuts1})--(\ref{eq:vbf-cuts4}) the LO curve exceeds the corresponding NLO-EW result, especially at large values of $m_{jj}$ where Sudakov suppression effects are becoming important. A similar trend can be observed in other distributions, such as the transverse momentum of the hardest tagging jet. The QED  shower has little impact on the NLO-EW results. 
%
%
%%%%
\begin{figure}
\includegraphics[angle=0,scale=0.5]{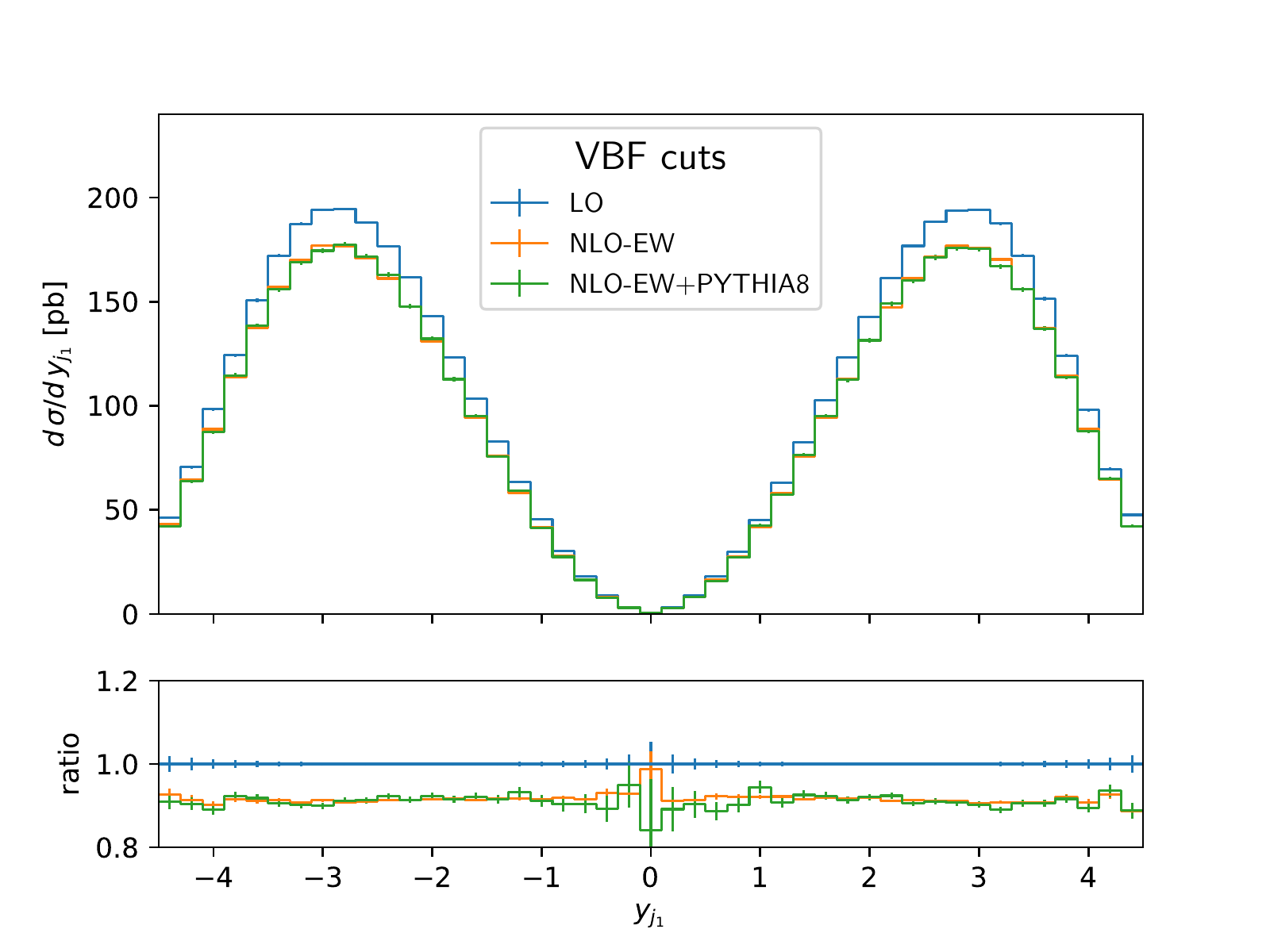}
\includegraphics[angle=0,scale=0.5]{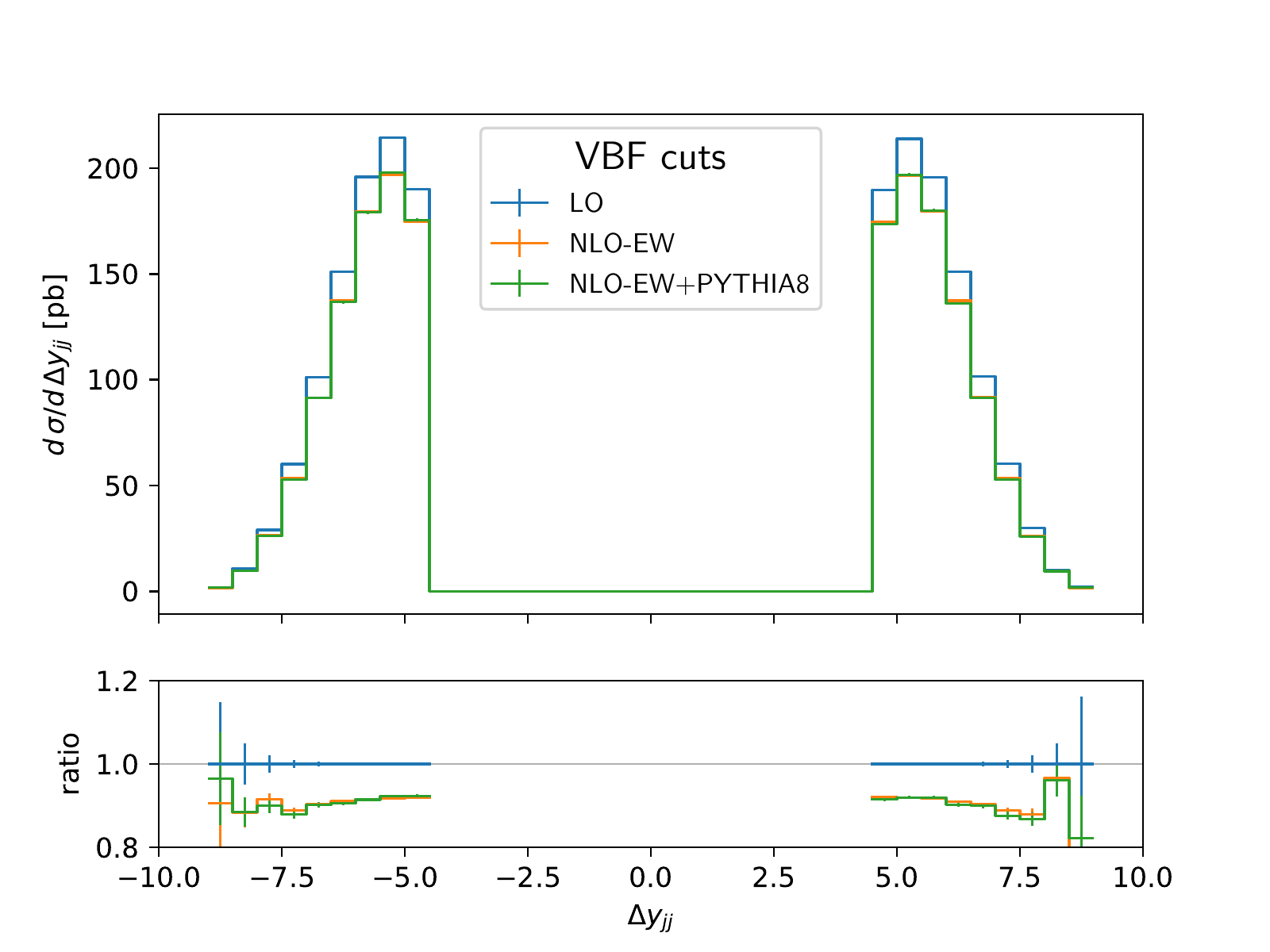}
\caption{\it 
\label{fig:ew-vbf-y}
Rapidity distribution of the hardest tagging jet (left) and rapidity separation of the two tagging jets (right) 
at LO (blue), NLO-EW (red), and NLO-EW+PS (green) accuracy within the VBF cuts of eqs.~(\ref{eq:vbf-cuts1})--(\ref{eq:vbf-cuts4}). 
The ratios of the NLO-EW+PS and the NLO-EW to the LO results are shown in the respective lower panels. 
}
\end{figure}
%%%%
% 
Also for the rapidity distributions of the tagging jets and their separation, shown in figure~\ref{fig:ew-vbf-y}, we barely observe differences between the NLO-EW and NLO-EW+PS results. 

\begin{figure}
\includegraphics[angle=0,scale=0.5]{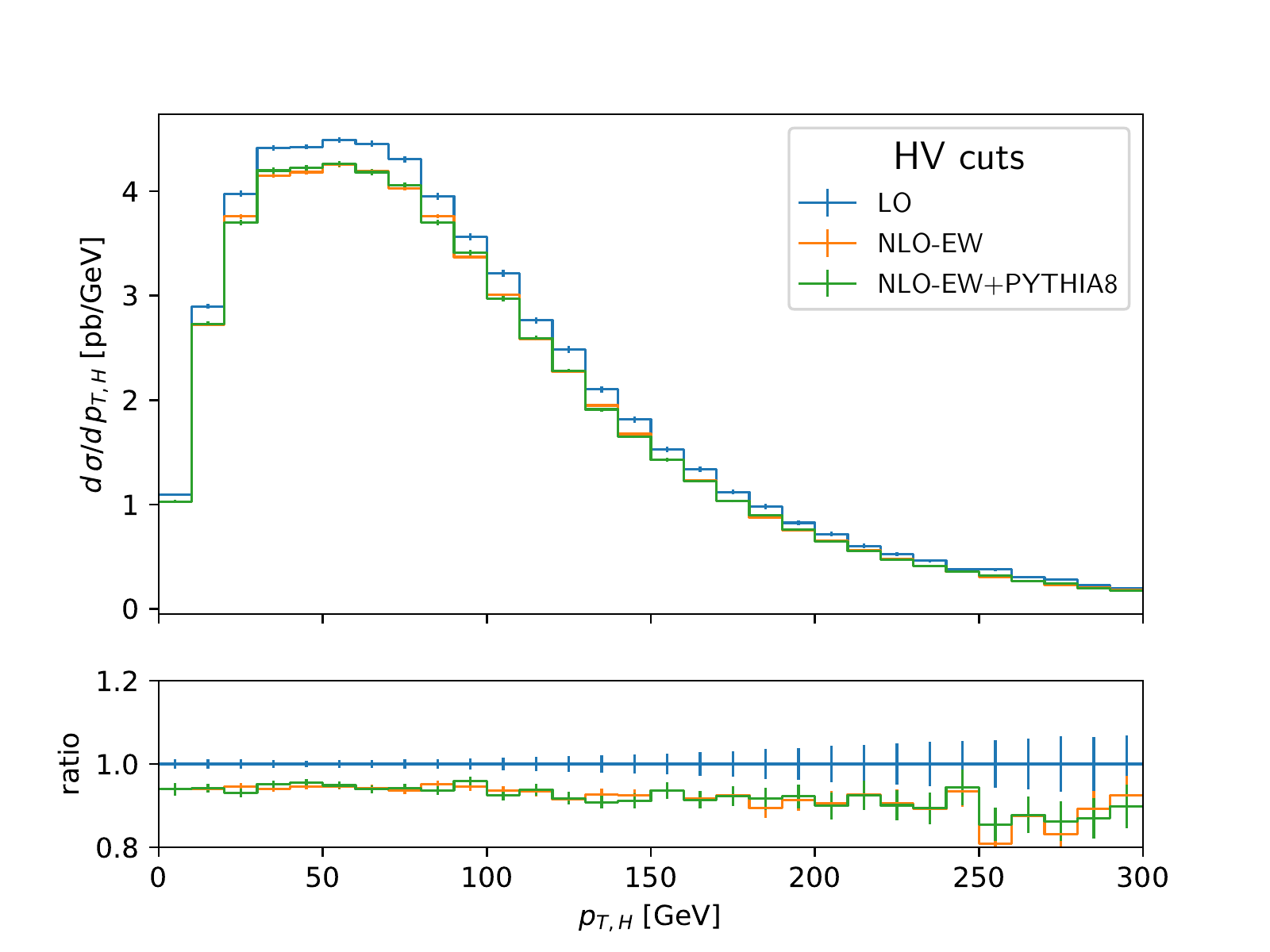}
\includegraphics[angle=0,scale=0.5]{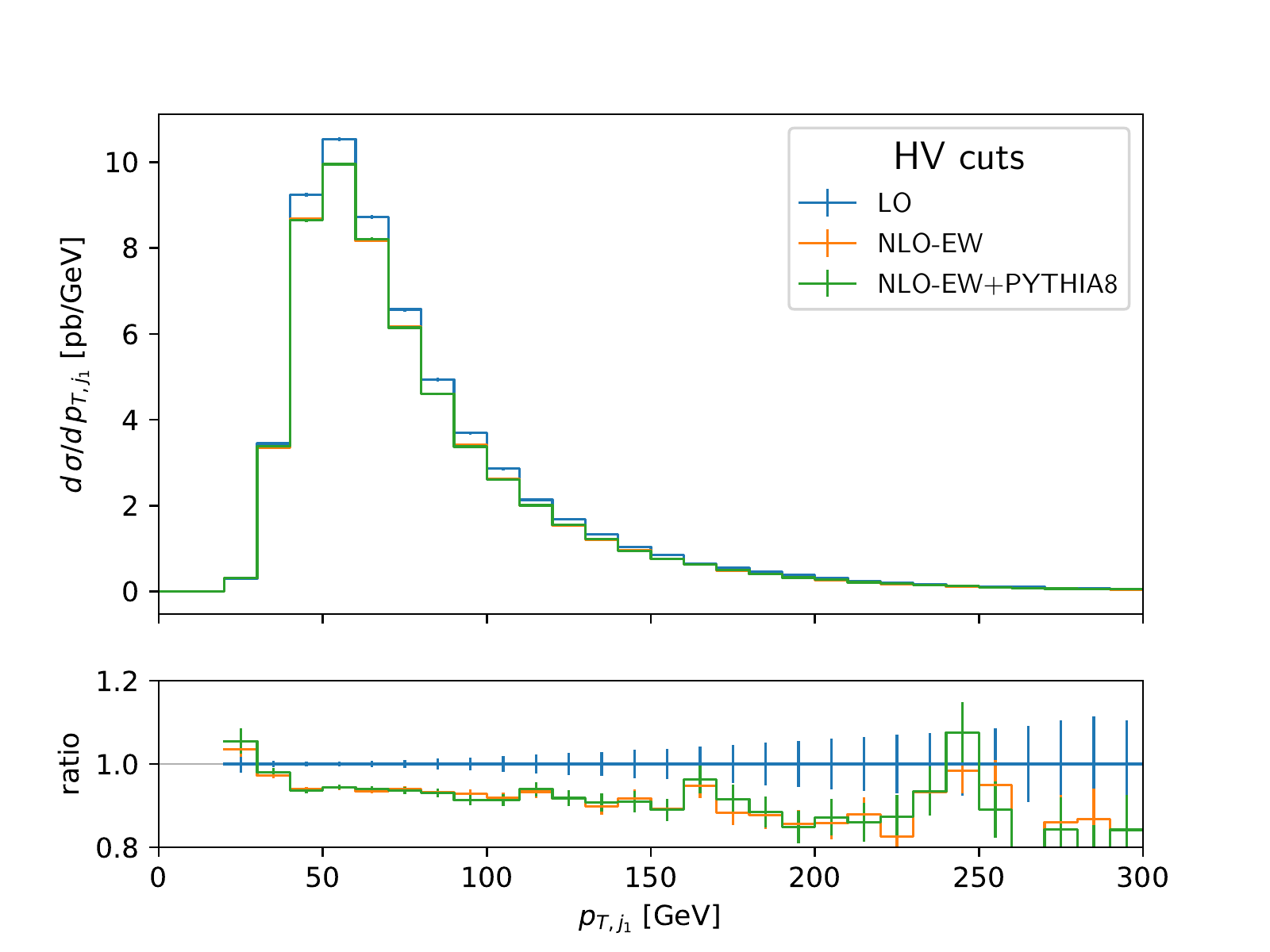}
\caption{\it 
\label{fig:ew-hv-pt}
Transverse-momentum distribution of the Higgs boson (left) and transverse momentum of the hardest tagging jet (right) 
at LO (blue), NLO-EW (red), and NLO-EW+PS (green) accuracy within the HV cuts of eqs.~(\ref{eq:hv-cuts1})--(\ref{eq:hv-cuts2}). 
The ratios of the NLO-EW+PS and the NLO-EW to the LO results are shown in the respective lower panels. 
}
\end{figure}

\begin{figure}
\includegraphics[angle=0,scale=0.5]{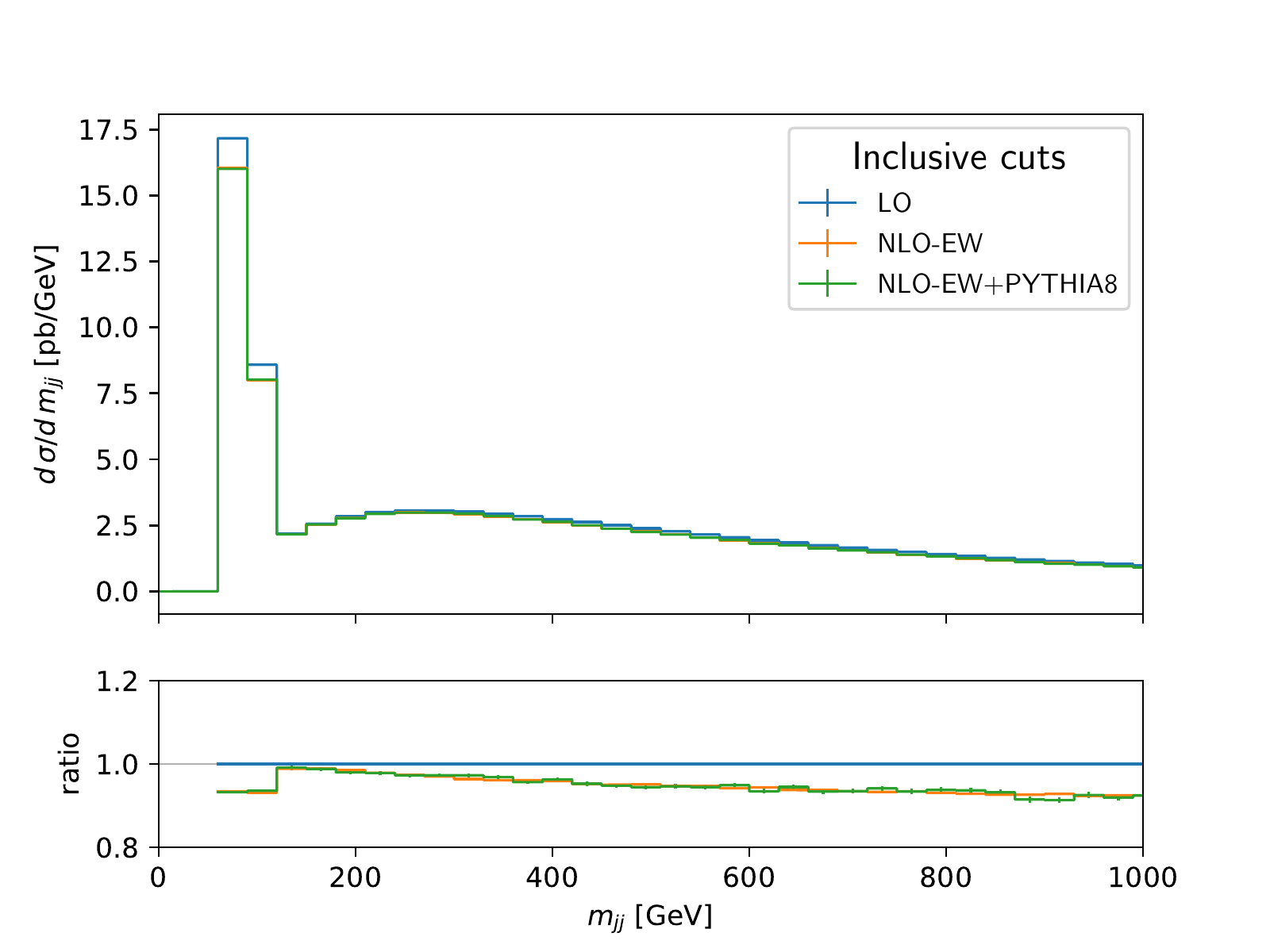}
\includegraphics[angle=0,scale=0.5]{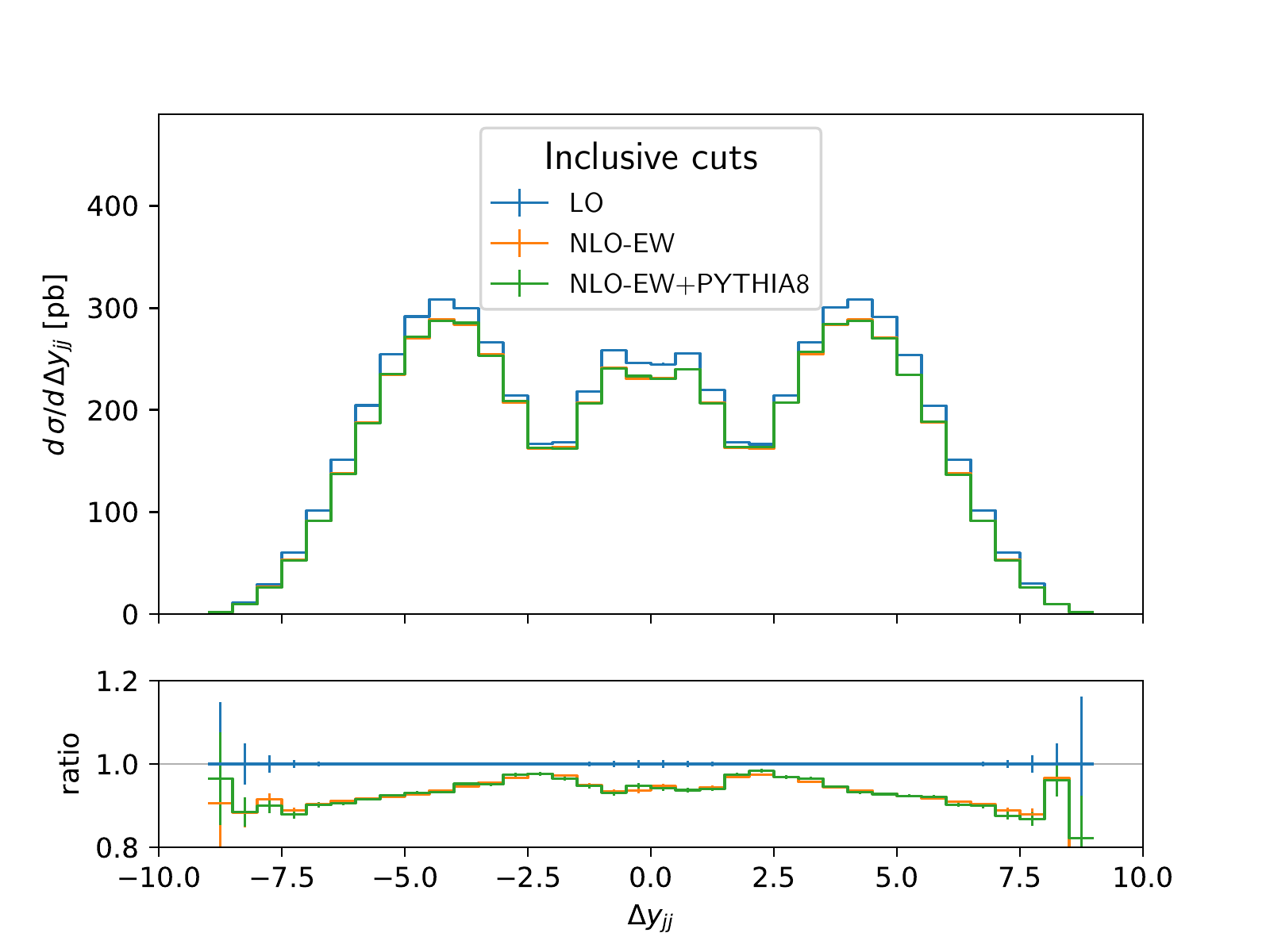}
	\caption{\it 
		\label{fig:ew-inc-jj}
		Invariant-mass distribution of the two tagging jets (left) and their rapidity separation (right) 
		at LO (blue), NLO-EW (red), and NLO-EW+PS (green) accuracy within the inclusive cuts of eqs.~(\ref{eq:inc-cuts1})--(\ref{eq:inc-cuts2}). The ratios of the NLO-EW+PS and the NLO-EW to the LO results are shown in the respective lower panels. 		}
\end{figure}

Predictions within the $HV$ setup follow a similar pattern: NLO-EW corrections typically yield negative contributions to the tails of moment-dependent distributions. QED shower effects barely change the NLO-EW results, as illustrated by figure~\ref{fig:ew-hv-pt} for the transverse-momentum distribution of the Higgs boson and the transverse momentum distribution of the hardest tagging jet. 
%
%
%%%%
%%%%
% 
%	
A similar trend can be observed in distributions within the inclusive cut setup. For instance, for the invariant-mass distribution of the two tagging jets and their rapidity separation, depicted in figure~\ref{fig:ew-inc-jj},  NLO-EW corrections shift the LO results to slightly smaller values. This effect is most pronounced at large values of $m_{jj}$. The QED shower has very little impact in any region of phase space. 
%
%%%%
%%%%
%	

%=================================================
%
\section{Conclusions and outlook}
\label{sec:conclusions}
In this work, we presented an implementation of the full EW production process of the \hjj{} final state in proton-proton collisions including NLO-QCD and NLO-EW corrections and their matching to QCD and QED showers within the \POWHEGBOX{} framework. Our code constitutes the first public implementation of this process including both $HV$ and VBF topologies as well as their interference at NLO-EW+PS level. At NLO-QCD+PS precision our results are consistent with  previous calculations for the individual $HV$ and VBF contributions. We confirm existing recommendations for using the dipole shower in \PYTHIAE{} for VBF-induced Higgs production.  The versatility of our code paves the way for comprehensive studies of PS generators and their optimal settings for \hjj{} production in more general setups.

In our sample phenomenological analyses we found that the NLO-EW corrections to \hjj{} production can be as large as the NLO-QCD contributions or even exceed them, depending on the considered selection cuts. EW corrections are typically negative. The influence of the QED shower on cross sections and differential distributions is only mild.

A natural extension of our work would be constituted by a combination of the NLO-QCD and EW corrections with QCD and QED showers.  While attractive from a user's point of view, adding such a feature to our code would require a technically non-trivial extension of the existing implementation that we leave for future work. 

%=================================================

%=================================================
%
\acknowledgments
%
%================================================= 
We are grateful for valuable discussions to Margherita Ghezzi, Alexander Karlberg, Seth Quackenbusch  and Laura Reina. 
This work has been supported by the German Research Foundation (DFG),  grant no.\ JA~1954/2-1. 
J.~S.\ has gratefully received financial support by the German Academic Scholarship Foundation for his research. The authors acknowledge support by the state of Baden-W\"urttemberg through bwHPC and the DFG  through grant no.~INST 39/963-1 FUGG.

%=================================================

\bibliographystyle{JHEP}
\bibliography{hjj}

\providecommand{\href}[2]{#2}\begingroup\raggedright\begin{thebibliography}{10}

\bibitem{Chatrchyan:2012xdj}
{\scshape CMS} collaboration, S.~Chatrchyan et~al., \emph{{Observation of a New
  Boson at a Mass of 125 GeV with the CMS Experiment at the LHC}},
  \href{https://doi.org/10.1016/j.physletb.2012.08.021}{\emph{Phys. Lett.}
  {\bfseries B716} (2012) 30}
  [\href{https://arxiv.org/abs/1207.7235}{{\ttfamily 1207.7235}}].

\bibitem{Aad:2012tfa}
{\scshape ATLAS Collaboration} collaboration, G.~Aad et~al., \emph{{Observation
  of a new particle in the search for the Standard Model Higgs boson with the
  ATLAS detector at the LHC}},
  \href{https://doi.org/10.1016/j.physletb.2012.08.020}{\emph{Phys.Lett.}
  {\bfseries B716} (2012) 1} [\href{https://arxiv.org/abs/1207.7214}{{\ttfamily
  1207.7214}}].

\bibitem{Rainwater:1999sd}
D.~L. Rainwater and D.~Zeppenfeld, \emph{{Observing $H\to W^*W^* \to e^\pm
  \mu\mp \not{p}_T$ in weak boson fusion with dual forward jet tagging at the
  CERN LHC}}, \href{https://doi.org/10.1103/PhysRevD.60.113004}{\emph{Phys.
  Rev. D} {\bfseries 60} (1999) 113004}
  [\href{https://arxiv.org/abs/hep-ph/9906218}{{\ttfamily hep-ph/9906218}}].

\bibitem{Ciccolini:2007ec}
M.~Ciccolini, A.~Denner and S.~Dittmaier, \emph{{Electroweak and QCD
  corrections to Higgs production via vector-boson fusion at the LHC}},
  \href{https://doi.org/10.1103/PhysRevD.77.013002}{\emph{Phys. Rev. D}
  {\bfseries 77} (2008) 013002}
  [\href{https://arxiv.org/abs/0710.4749}{{\ttfamily 0710.4749}}].

\bibitem{Han:1992hr}
T.~Han, G.~Valencia and S.~Willenbrock, \emph{{Structure function approach to
  vector boson scattering in p p collisions}},
  \href{https://doi.org/10.1103/PhysRevLett.69.3274}{\emph{Phys. Rev. Lett.}
  {\bfseries 69} (1992) 3274}
  [\href{https://arxiv.org/abs/hep-ph/9206246}{{\ttfamily hep-ph/9206246}}].

\bibitem{Figy:2003nv}
T.~Figy, C.~Oleari and D.~Zeppenfeld, \emph{{Next-to-leading order jet
  distributions for Higgs boson production via weak boson fusion}},
  \href{https://doi.org/10.1103/PhysRevD.68.073005}{\emph{Phys. Rev. D}
  {\bfseries 68} (2003) 073005}
  [\href{https://arxiv.org/abs/hep-ph/0306109}{{\ttfamily hep-ph/0306109}}].

\bibitem{Berger:2004pca}
E.~L. Berger and J.~M. Campbell, \emph{{Higgs boson production in weak boson
  fusion at next-to-leading order}},
  \href{https://doi.org/10.1103/PhysRevD.70.073011}{\emph{Phys. Rev. D}
  {\bfseries 70} (2004) 073011}
  [\href{https://arxiv.org/abs/hep-ph/0403194}{{\ttfamily hep-ph/0403194}}].

\bibitem{Arnold:2008rz}
K.~Arnold et~al., \emph{{VBFNLO: A Parton level Monte Carlo for processes with
  electroweak bosons}},
  \href{https://doi.org/10.1016/j.cpc.2009.03.006}{\emph{Comput. Phys. Commun.}
  {\bfseries 180} (2009) 1661}
  [\href{https://arxiv.org/abs/0811.4559}{{\ttfamily 0811.4559}}].

\bibitem{Campbell:2002tg}
J.~M. Campbell and R.~K. Ellis, \emph{{Next-to-Leading Order Corrections to
  $W^+$ 2 jet and $Z^+$ 2 Jet Production at Hadron Colliders}},
  \href{https://doi.org/10.1103/PhysRevD.65.113007}{\emph{Phys. Rev. D}
  {\bfseries 65} (2002) 113007}
  [\href{https://arxiv.org/abs/hep-ph/0202176}{{\ttfamily hep-ph/0202176}}].

\bibitem{Figy:2007kv}
T.~Figy, V.~Hankele and D.~Zeppenfeld, \emph{{Next-to-leading order QCD
  corrections to Higgs plus three jet production in vector-boson fusion}},
  \href{https://doi.org/10.1088/1126-6708/2008/02/076}{\emph{JHEP} {\bfseries
  02} (2008) 076} [\href{https://arxiv.org/abs/0710.5621}{{\ttfamily
  0710.5621}}].

\bibitem{Bolzoni:2010xr}
P.~Bolzoni, F.~Maltoni, S.-O. Moch and M.~Zaro, \emph{{Higgs production via
  vector-boson fusion at NNLO in QCD}},
  \href{https://doi.org/10.1103/PhysRevLett.105.011801}{\emph{Phys. Rev. Lett.}
  {\bfseries 105} (2010) 011801}
  [\href{https://arxiv.org/abs/1003.4451}{{\ttfamily 1003.4451}}].

\bibitem{Bolzoni:2011cu}
P.~Bolzoni, F.~Maltoni, S.-O. Moch and M.~Zaro, \emph{{Vector boson fusion at
  NNLO in QCD: SM Higgs and beyond}},
  \href{https://doi.org/10.1103/PhysRevD.85.035002}{\emph{Phys. Rev. D}
  {\bfseries 85} (2012) 035002}
  [\href{https://arxiv.org/abs/1109.3717}{{\ttfamily 1109.3717}}].

\bibitem{Cacciari:2015jma}
M.~Cacciari, F.~A. Dreyer, A.~Karlberg, G.~P. Salam and G.~Zanderighi,
  \emph{{Fully Differential Vector-Boson-Fusion Higgs Production at
  Next-to-Next-to-Leading Order}},
  \href{https://doi.org/10.1103/PhysRevLett.115.082002}{\emph{Phys. Rev. Lett.}
  {\bfseries 115} (2015) 082002}
  [\href{https://arxiv.org/abs/1506.02660}{{\ttfamily 1506.02660}}].

\bibitem{Dreyer:2016oyx}
F.~A. Dreyer and A.~Karlberg, \emph{{Vector-Boson Fusion Higgs Production at
  Three Loops in QCD}},
  \href{https://doi.org/10.1103/PhysRevLett.117.072001}{\emph{Phys. Rev. Lett.}
  {\bfseries 117} (2016) 072001}
  [\href{https://arxiv.org/abs/1606.00840}{{\ttfamily 1606.00840}}].

\bibitem{Cruz-Martinez:2018rod}
J.~Cruz-Martinez, T.~Gehrmann, E.~W.~N. Glover and A.~Huss, \emph{{Second-order
  QCD effects in Higgs boson production through vector boson fusion}},
  \href{https://doi.org/10.1016/j.physletb.2018.04.046}{\emph{Phys. Lett. B}
  {\bfseries 781} (2018) 672}
  [\href{https://arxiv.org/abs/1802.02445}{{\ttfamily 1802.02445}}].

\bibitem{Buckley:2021gfw}
A.~Buckley et~al., \emph{{A comparative study of Higgs boson production from
  vector-boson fusion}},
  \href{https://doi.org/10.1007/JHEP11(2021)108}{\emph{JHEP} {\bfseries 11}
  (2021) 108} [\href{https://arxiv.org/abs/2105.11399}{{\ttfamily
  2105.11399}}].

\bibitem{Chen:2021phj}
T.~Chen, T.~M. Figy and S.~Pl\"atzer, \emph{{NLO Multijet Merging for Higgs
  Production Beyond the VBF Approximation}},
  \href{https://arxiv.org/abs/2109.03730}{{\ttfamily 2109.03730}}.

\bibitem{Liu:2019tuy}
T.~Liu, K.~Melnikov and A.~A. Penin, \emph{{Nonfactorizable QCD Effects in
  Higgs Boson Production via Vector Boson Fusion}},
  \href{https://doi.org/10.1103/PhysRevLett.123.122002}{\emph{Phys. Rev. Lett.}
  {\bfseries 123} (2019) 122002}
  [\href{https://arxiv.org/abs/1906.10899}{{\ttfamily 1906.10899}}].

\bibitem{Dreyer:2020urf}
F.~A. Dreyer, A.~Karlberg and L.~Tancredi, \emph{{On the impact of
  non-factorisable corrections in VBF single and double Higgs production}},
  \href{https://doi.org/10.1007/JHEP10(2020)131}{\emph{JHEP} {\bfseries 10}
  (2020) 131} [\href{https://arxiv.org/abs/2005.11334}{{\ttfamily
  2005.11334}}].

\bibitem{Bittrich:2021ztq}
C.~Bittrich, P.~Kirchgae\ss{}er, A.~Papaefstathiou, S.~Pl\"atzer and S.~Todt,
  \emph{{Soft QCD Effects in VBS/VBF Topologies}},
  \href{https://arxiv.org/abs/2110.01623}{{\ttfamily 2110.01623}}.

\bibitem{Brein:2003wg}
O.~Brein, A.~Djouadi and R.~Harlander, \emph{{NNLO QCD corrections to the
  Higgs-strahlung processes at hadron colliders}},
  \href{https://doi.org/10.1016/j.physletb.2003.10.112}{\emph{Phys. Lett. B}
  {\bfseries 579} (2004) 149}
  [\href{https://arxiv.org/abs/hep-ph/0307206}{{\ttfamily hep-ph/0307206}}].

\bibitem{Brein:2011vx}
O.~Brein, R.~Harlander, M.~Wiesemann and T.~Zirke, \emph{{Top-Quark Mediated
  Effects in Hadronic Higgs-Strahlung}},
  \href{https://doi.org/10.1140/epjc/s10052-012-1868-6}{\emph{Eur. Phys. J. C}
  {\bfseries 72} (2012) 1868}
  [\href{https://arxiv.org/abs/1111.0761}{{\ttfamily 1111.0761}}].

\bibitem{Brein:2012ne}
O.~Brein, R.~V. Harlander and T.~J.~E. Zirke, \emph{{vh@nnlo - Higgs Strahlung
  at hadron colliders}},
  \href{https://doi.org/10.1016/j.cpc.2012.11.002}{\emph{Comput. Phys. Commun.}
  {\bfseries 184} (2013) 998}
  [\href{https://arxiv.org/abs/1210.5347}{{\ttfamily 1210.5347}}].

\bibitem{PhysRevD.42.2253}
B.~A. Kniehl, \emph{Associated production of higgs and $z$ bosons from gluon
  fusion in hadron collisions},
  \href{https://doi.org/10.1103/PhysRevD.42.2253}{\emph{Phys. Rev. D}
  {\bfseries 42} (1990) 2253}.

\bibitem{Altenkamp:2012sx}
L.~Altenkamp, S.~Dittmaier, R.~V. Harlander, H.~Rzehak and T.~J.~E. Zirke,
  \emph{{Gluon-induced Higgs-strahlung at next-to-leading order QCD}},
  \href{https://doi.org/10.1007/JHEP02(2013)078}{\emph{JHEP} {\bfseries 02}
  (2013) 078} [\href{https://arxiv.org/abs/1211.5015}{{\ttfamily 1211.5015}}].

\bibitem{Ferrera:2011bk}
G.~Ferrera, M.~Grazzini and F.~Tramontano, \emph{{Associated WH production at
  hadron colliders: a fully exclusive QCD calculation at NNLO}},
  \href{https://doi.org/10.1103/PhysRevLett.107.152003}{\emph{Phys. Rev. Lett.}
  {\bfseries 107} (2011) 152003}
  [\href{https://arxiv.org/abs/1107.1164}{{\ttfamily 1107.1164}}].

\bibitem{Ferrera:2014lca}
G.~Ferrera, M.~Grazzini and F.~Tramontano, \emph{{Associated ZH production at
  hadron colliders: the fully differential NNLO QCD calculation}},
  \href{https://doi.org/10.1016/j.physletb.2014.11.040}{\emph{Phys. Lett. B}
  {\bfseries 740} (2015) 51} [\href{https://arxiv.org/abs/1407.4747}{{\ttfamily
  1407.4747}}].

\bibitem{Campbell:2016jau}
J.~M. Campbell, R.~K. Ellis and C.~Williams, \emph{{Associated production of a
  Higgs boson at NNLO}},
  \href{https://doi.org/10.1007/JHEP06(2016)179}{\emph{JHEP} {\bfseries 06}
  (2016) 179} [\href{https://arxiv.org/abs/1601.00658}{{\ttfamily
  1601.00658}}].

\bibitem{Ferrera:2017zex}
G.~Ferrera, G.~Somogyi and F.~Tramontano, \emph{{Associated production of a
  Higgs boson decaying into bottom quarks at the LHC in full NNLO QCD}},
  \href{https://doi.org/10.1016/j.physletb.2018.03.021}{\emph{Phys. Lett. B}
  {\bfseries 780} (2018) 346}
  [\href{https://arxiv.org/abs/1705.10304}{{\ttfamily 1705.10304}}].

\bibitem{Ciccolini:2003jy}
M.~L. Ciccolini, S.~Dittmaier and M.~Kramer, \emph{{Electroweak radiative
  corrections to associated WH and ZH production at hadron colliders}},
  \href{https://doi.org/10.1103/PhysRevD.68.073003}{\emph{Phys. Rev. D}
  {\bfseries 68} (2003) 073003}
  [\href{https://arxiv.org/abs/hep-ph/0306234}{{\ttfamily hep-ph/0306234}}].

\bibitem{Denner:2011id}
A.~Denner, S.~Dittmaier, S.~Kallweit and A.~Muck, \emph{{Electroweak
  corrections to Higgs-strahlung off W/Z bosons at the Tevatron and the LHC
  with HAWK}}, \href{https://doi.org/10.1007/JHEP03(2012)075}{\emph{JHEP}
  {\bfseries 03} (2012) 075} [\href{https://arxiv.org/abs/1112.5142}{{\ttfamily
  1112.5142}}].

\bibitem{Ciccolini:2007jr}
M.~Ciccolini, A.~Denner and S.~Dittmaier, \emph{{Strong and electroweak
  corrections to the production of Higgs + 2jets via weak interactions at the
  LHC}}, \href{https://doi.org/10.1103/PhysRevLett.99.161803}{\emph{Phys. Rev.
  Lett.} {\bfseries 99} (2007) 161803}
  [\href{https://arxiv.org/abs/0707.0381}{{\ttfamily 0707.0381}}].

\bibitem{Figy:2010ct}
T.~Figy, S.~Palmer and G.~Weiglein, \emph{{Higgs Production via Weak Boson
  Fusion in the Standard Model and the MSSM}},
  \href{https://doi.org/10.1007/JHEP02(2012)105}{\emph{JHEP} {\bfseries 02}
  (2012) 105} [\href{https://arxiv.org/abs/1012.4789}{{\ttfamily 1012.4789}}].

\bibitem{Denner:2014cla}
A.~Denner, S.~Dittmaier, S.~Kallweit and A.~M\"uck, \emph{{HAWK 2.0: A Monte
  Carlo program for Higgs production in vector-boson fusion and Higgs strahlung
  at hadron colliders}},
  \href{https://doi.org/10.1016/j.cpc.2015.04.021}{\emph{Comput. Phys. Commun.}
  {\bfseries 195} (2015) 161}
  [\href{https://arxiv.org/abs/1412.5390}{{\ttfamily 1412.5390}}].

\bibitem{Denner:2000jv}
A.~Denner and S.~Pozzorini, \emph{{One loop leading logarithms in electroweak
  radiative corrections. 1. Results}},
  \href{https://doi.org/10.1007/s100520100551}{\emph{Eur. Phys. J. C}
  {\bfseries 18} (2001) 461}
  [\href{https://arxiv.org/abs/hep-ph/0010201}{{\ttfamily hep-ph/0010201}}].

\bibitem{Sjostrand:2006za}
T.~Sjostrand, S.~Mrenna and P.~Z. Skands, \emph{{PYTHIA 6.4 Physics and
  Manual}}, \href{https://doi.org/10.1088/1126-6708/2006/05/026}{\emph{JHEP}
  {\bfseries 05} (2006) 026}
  [\href{https://arxiv.org/abs/hep-ph/0603175}{{\ttfamily hep-ph/0603175}}].

\bibitem{Sjostrand:2014zea}
T.~Sj\"ostrand, S.~Ask, J.~R. Christiansen, R.~Corke, N.~Desai, P.~Ilten
  et~al., \emph{{An introduction to PYTHIA 8.2}},
  \href{https://doi.org/10.1016/j.cpc.2015.01.024}{\emph{Comput. Phys. Commun.}
  {\bfseries 191} (2015) 159}
  [\href{https://arxiv.org/abs/1410.3012}{{\ttfamily 1410.3012}}].

\bibitem{Corcella:2000bw}
G.~Corcella, I.~G. Knowles, G.~Marchesini, S.~Moretti, K.~Odagiri,
  P.~Richardson et~al., \emph{{HERWIG 6: An Event generator for hadron emission
  reactions with interfering gluons (including supersymmetric processes)}},
  \href{https://doi.org/10.1088/1126-6708/2001/01/010}{\emph{JHEP} {\bfseries
  01} (2001) 010} [\href{https://arxiv.org/abs/hep-ph/0011363}{{\ttfamily
  hep-ph/0011363}}].

\bibitem{Bellm:2015jjp}
J.~Bellm et~al., \emph{{Herwig 7.0/Herwig++ 3.0 release note}},
  \href{https://doi.org/10.1140/epjc/s10052-016-4018-8}{\emph{Eur. Phys. J. C}
  {\bfseries 76} (2016) 196}
  [\href{https://arxiv.org/abs/1512.01178}{{\ttfamily 1512.01178}}].

\bibitem{Gleisberg:2008ta}
T.~Gleisberg, S.~Hoeche, F.~Krauss, M.~Schonherr, S.~Schumann, F.~Siegert
  et~al., \emph{{Event generation with SHERPA 1.1}},
  \href{https://doi.org/10.1088/1126-6708/2009/02/007}{\emph{JHEP} {\bfseries
  02} (2009) 007} [\href{https://arxiv.org/abs/0811.4622}{{\ttfamily
  0811.4622}}].

\bibitem{Frixione:2002ik}
S.~Frixione and B.~R. Webber, \emph{{Matching NLO QCD computations and parton
  shower simulations}},
  \href{https://doi.org/10.1088/1126-6708/2002/06/029}{\emph{JHEP} {\bfseries
  06} (2002) 029} [\href{https://arxiv.org/abs/hep-ph/0204244}{{\ttfamily
  hep-ph/0204244}}].

\bibitem{Nason:2004rx}
P.~Nason, \emph{{A New method for combining NLO QCD with shower Monte Carlo
  algorithms}},
  \href{https://doi.org/10.1088/1126-6708/2004/11/040}{\emph{JHEP} {\bfseries
  11} (2004) 040} [\href{https://arxiv.org/abs/hep-ph/0409146}{{\ttfamily
  hep-ph/0409146}}].

\bibitem{Frixione:2007vw}
S.~Frixione, P.~Nason and C.~Oleari, \emph{{Matching NLO QCD computations with
  Parton Shower simulations: the POWHEG method}},
  \href{https://doi.org/10.1088/1126-6708/2007/11/070}{\emph{JHEP} {\bfseries
  11} (2007) 070} [\href{https://arxiv.org/abs/0709.2092}{{\ttfamily
  0709.2092}}].

\bibitem{Alioli:2010xd}
S.~Alioli, P.~Nason, C.~Oleari and E.~Re, \emph{{A general framework for
  implementing NLO calculations in shower Monte Carlo programs: the POWHEG
  BOX}}, \href{https://doi.org/10.1007/JHEP06(2010)043}{\emph{JHEP} {\bfseries
  06} (2010) 043} [\href{https://arxiv.org/abs/1002.2581}{{\ttfamily
  1002.2581}}].

\bibitem{Nason:2009ai}
P.~Nason and C.~Oleari, \emph{{NLO Higgs boson production via vector-boson
  fusion matched with shower in POWHEG}},
  \href{https://doi.org/10.1007/JHEP02(2010)037}{\emph{JHEP} {\bfseries 02}
  (2010) 037} [\href{https://arxiv.org/abs/0911.5299}{{\ttfamily 0911.5299}}].

\bibitem{Frixione:2013mta}
S.~Frixione, P.~Torrielli and M.~Zaro, \emph{{Higgs production through
  vector-boson fusion at the NLO matched with parton showers}},
  \href{https://doi.org/10.1016/j.physletb.2013.08.030}{\emph{Phys. Lett. B}
  {\bfseries 726} (2013) 273}
  [\href{https://arxiv.org/abs/1304.7927}{{\ttfamily 1304.7927}}].

\bibitem{Campanario:2013fsa}
F.~Campanario, T.~M. Figy, S.~Pl\"atzer and M.~Sj\"odahl, \emph{{Electroweak
  Higgs Boson Plus Three Jet Production at Next-to-Leading-Order QCD}},
  \href{https://doi.org/10.1103/PhysRevLett.111.211802}{\emph{Phys. Rev. Lett.}
  {\bfseries 111} (2013) 211802}
  [\href{https://arxiv.org/abs/1308.2932}{{\ttfamily 1308.2932}}].

\bibitem{Alwall:2014hca}
J.~Alwall, R.~Frederix, S.~Frixione, V.~Hirschi, F.~Maltoni, O.~Mattelaer
  et~al., \emph{{The automated computation of tree-level and next-to-leading
  order differential cross sections, and their matching to parton shower
  simulations}}, \href{https://doi.org/10.1007/JHEP07(2014)079}{\emph{JHEP}
  {\bfseries 07} (2014) 079} [\href{https://arxiv.org/abs/1405.0301}{{\ttfamily
  1405.0301}}].

\bibitem{Jager:2020hkz}
B.~J\"ager, A.~Karlberg, S.~Pl\"atzer, J.~Scheller and M.~Zaro,
  \emph{{Parton-shower effects in Higgs production via Vector-Boson Fusion}},
  \href{https://doi.org/10.1140/epjc/s10052-020-8326-7}{\emph{Eur. Phys. J. C}
  {\bfseries 80} (2020) 756}
  [\href{https://arxiv.org/abs/2003.12435}{{\ttfamily 2003.12435}}].

\bibitem{Luisoni:2013cuh}
G.~Luisoni, P.~Nason, C.~Oleari and F.~Tramontano, \emph{{$HW^{\pm}$/HZ + 0 and
  1 jet at NLO with the POWHEG BOX interfaced to GoSam and their merging within
  MiNLO}}, \href{https://doi.org/10.1007/JHEP10(2013)083}{\emph{JHEP}
  {\bfseries 10} (2013) 083} [\href{https://arxiv.org/abs/1306.2542}{{\ttfamily
  1306.2542}}].

\bibitem{Granata:2017iod}
F.~Granata, J.~M. Lindert, C.~Oleari and S.~Pozzorini, \emph{{NLO QCD+EW
  predictions for HV and HV +jet production including parton-shower effects}},
  \href{https://doi.org/10.1007/JHEP09(2017)012}{\emph{JHEP} {\bfseries 09}
  (2017) 012} [\href{https://arxiv.org/abs/1706.03522}{{\ttfamily
  1706.03522}}].

\bibitem{Jezo:2015aia}
T.~Je\v{z}o and P.~Nason, \emph{{On the Treatment of Resonances in
  Next-to-Leading Order Calculations Matched to a Parton Shower}},
  \href{https://doi.org/10.1007/JHEP12(2015)065}{\emph{JHEP} {\bfseries 12}
  (2015) 065} [\href{https://arxiv.org/abs/1509.09071}{{\ttfamily
  1509.09071}}].

\bibitem{Frixione:1995ms}
S.~Frixione, Z.~Kunszt and A.~Signer, \emph{{Three jet cross-sections to
  next-to-leading order}},
  \href{https://doi.org/10.1016/0550-3213(96)00110-1}{\emph{Nucl. Phys.}
  {\bfseries B467} (1996) 399}
  [\href{https://arxiv.org/abs/hep-ph/9512328}{{\ttfamily hep-ph/9512328}}].

\bibitem{Barze:2012tt}
L.~Barze, G.~Montagna, P.~Nason, O.~Nicrosini and F.~Piccinini,
  \emph{{Implementation of electroweak corrections in the POWHEG BOX: single W
  production}}, \href{https://doi.org/10.1007/JHEP04(2012)037}{\emph{JHEP}
  {\bfseries 04} (2012) 037} [\href{https://arxiv.org/abs/1202.0465}{{\ttfamily
  1202.0465}}].

\bibitem{Barze:2013fru}
L.~Barze, G.~Montagna, P.~Nason, O.~Nicrosini, F.~Piccinini and A.~Vicini,
  \emph{{Neutral current Drell-Yan with combined QCD and electroweak
  corrections in the POWHEG BOX}},
  \href{https://doi.org/10.1140/epjc/s10052-013-2474-y}{\emph{Eur. Phys. J. C}
  {\bfseries 73} (2013) 2474}
  [\href{https://arxiv.org/abs/1302.4606}{{\ttfamily 1302.4606}}].

\bibitem{Actis:2016mpe}
S.~Actis, A.~Denner, L.~Hofer, J.-N. Lang, A.~Scharf and S.~Uccirati,
  \emph{{RECOLA: REcursive Computation of One-Loop Amplitudes}},
  \href{https://doi.org/10.1016/j.cpc.2017.01.004}{\emph{Comput. Phys. Commun.}
  {\bfseries 214} (2017) 140}
  [\href{https://arxiv.org/abs/1605.01090}{{\ttfamily 1605.01090}}].

\bibitem{Denner:2017vms}
A.~Denner, J.-N. Lang and S.~Uccirati, \emph{{NLO electroweak corrections in
  extended Higgs Sectors with RECOLA2}},
  \href{https://doi.org/10.1007/JHEP07(2017)087}{\emph{JHEP} {\bfseries 07}
  (2017) 087} [\href{https://arxiv.org/abs/1705.06053}{{\ttfamily
  1705.06053}}].

\bibitem{Denner:2017wsf}
A.~Denner, J.-N. Lang and S.~Uccirati, \emph{{Recola2: REcursive Computation of
  One-Loop Amplitudes 2}},
  \href{https://doi.org/10.1016/j.cpc.2017.11.013}{\emph{Comput. Phys. Commun.}
  {\bfseries 224} (2018) 346}
  [\href{https://arxiv.org/abs/1711.07388}{{\ttfamily 1711.07388}}].

\bibitem{Chiesa:2020ttl}
M.~Chiesa, C.~Oleari and E.~Re, \emph{{NLO QCD+NLO EW corrections to diboson
  production matched to parton shower}},
  \href{https://doi.org/10.1140/epjc/s10052-020-8419-3}{\emph{Eur. Phys. J. C}
  {\bfseries 80} (2020) 849}
  [\href{https://arxiv.org/abs/2005.12146}{{\ttfamily 2005.12146}}].

\bibitem{Chiesa:2019ulk}
M.~Chiesa, A.~Denner, J.-N. Lang and M.~Pellen, \emph{{An event generator for
  same-sign W-boson scattering at the LHC including electroweak corrections}},
  \href{https://doi.org/10.1140/epjc/s10052-019-7290-6}{\emph{Eur. Phys. J. C}
  {\bfseries 79} (2019) 788}
  [\href{https://arxiv.org/abs/1906.01863}{{\ttfamily 1906.01863}}].

\bibitem{Denner:2016kdg}
A.~Denner, S.~Dittmaier and L.~Hofer, \emph{{Collier: a fortran-based Complex
  One-Loop LIbrary in Extended Regularizations}},
  \href{https://doi.org/10.1016/j.cpc.2016.10.013}{\emph{Comput. Phys. Commun.}
  {\bfseries 212} (2017) 220}
  [\href{https://arxiv.org/abs/1604.06792}{{\ttfamily 1604.06792}}].

\bibitem{Alwall:2006yp}
J.~Alwall et~al., \emph{{A Standard format for Les Houches event files}},
  \href{https://doi.org/10.1016/j.cpc.2006.11.010}{\emph{Comput. Phys. Commun.}
  {\bfseries 176} (2007) 300}
  [\href{https://arxiv.org/abs/hep-ph/0609017}{{\ttfamily hep-ph/0609017}}].

\bibitem{Alwall:2007st}
J.~Alwall, P.~Demin, S.~de~Visscher, R.~Frederix, M.~Herquet, F.~Maltoni
  et~al., \emph{{MadGraph/MadEvent v4: The New Web Generation}},
  \href{https://doi.org/10.1088/1126-6708/2007/09/028}{\emph{JHEP} {\bfseries
  09} (2007) 028} [\href{https://arxiv.org/abs/0706.2334}{{\ttfamily
  0706.2334}}].

\bibitem{Denner:2018opp}
A.~Denner, S.~Dittmaier and J.-N. Lang, \emph{{Renormalization of mixing
  angles}}, \href{https://doi.org/10.1007/JHEP11(2018)104}{\emph{JHEP}
  {\bfseries 11} (2018) 104}
  [\href{https://arxiv.org/abs/1808.03466}{{\ttfamily 1808.03466}}].

\bibitem{Buckley:2014ana}
A.~Buckley, J.~Ferrando, S.~Lloyd, K.~Nordstr\"om, B.~Page, M.~R\"ufenacht
  et~al., \emph{{LHAPDF6: parton density access in the LHC precision era}},
  \href{https://doi.org/10.1140/epjc/s10052-015-3318-8}{\emph{Eur. Phys. J. C}
  {\bfseries 75} (2015) 132} [\href{https://arxiv.org/abs/1412.7420}{{\ttfamily
  1412.7420}}].

\bibitem{Cacciari:2008gp}
M.~Cacciari, G.~P. Salam and G.~Soyez, \emph{{The anti-$k_t$ jet clustering
  algorithm}}, \href{https://doi.org/10.1088/1126-6708/2008/04/063}{\emph{JHEP}
  {\bfseries 04} (2008) 063} [\href{https://arxiv.org/abs/0802.1189}{{\ttfamily
  0802.1189}}].

\bibitem{Cacciari:2011ma}
M.~Cacciari, G.~P. Salam and G.~Soyez, \emph{{FastJet User Manual}},
  \href{https://doi.org/10.1140/epjc/s10052-012-1896-2}{\emph{Eur. Phys. J.}
  {\bfseries C72} (2012) 1896}
  [\href{https://arxiv.org/abs/1111.6097}{{\ttfamily 1111.6097}}].

\bibitem{Zyla:2020zbs}
{\scshape Particle Data Group} collaboration, P.~Zyla et~al., \emph{{Review of
  Particle Physics}}, \href{https://doi.org/10.1093/ptep/ptaa104}{\emph{PTEP}
  {\bfseries 2020} (2020) 083C01}.

\bibitem{Skands:2014pea}
P.~Skands, S.~Carrazza and J.~Rojo, \emph{{Tuning PYTHIA 8.1: the Monash 2013
  Tune}}, \href{https://doi.org/10.1140/epjc/s10052-014-3024-y}{\emph{Eur.
  Phys. J. C} {\bfseries 74} (2014) 3024}
  [\href{https://arxiv.org/abs/1404.5630}{{\ttfamily 1404.5630}}].

\bibitem{DelDuca:2006hk}
V.~Del~Duca, G.~Klamke, D.~Zeppenfeld, M.~L. Mangano, M.~Moretti, F.~Piccinini
  et~al., \emph{{Monte Carlo studies of the jet activity in Higgs + 2 jet
  events}}, \href{https://doi.org/10.1088/1126-6708/2006/10/016}{\emph{JHEP}
  {\bfseries 10} (2006) 016}
  [\href{https://arxiv.org/abs/hep-ph/0608158}{{\ttfamily hep-ph/0608158}}].

\bibitem{Cabouat:2017rzi}
B.~Cabouat and T.~Sj\"ostrand, \emph{{Some Dipole Shower Studies}},
  \href{https://doi.org/10.1140/epjc/s10052-018-5645-z}{\emph{Eur. Phys. J. C}
  {\bfseries 78} (2018) 226}
  [\href{https://arxiv.org/abs/1710.00391}{{\ttfamily 1710.00391}}].

\bibitem{Konar:2022bgc}
P.~Konar and V.~S. Ngairangbam, \emph{{Influence of QCD parton showers in deep
  learning invisible Higgs bosons through vector boson fusion}},
  \href{https://doi.org/10.1103/PhysRevD.105.113003}{\emph{Phys. Rev. D}
  {\bfseries 105} (2022) 113003}
  [\href{https://arxiv.org/abs/2201.01040}{{\ttfamily 2201.01040}}].

\end{thebibliography}\endgroup

\end{document}